\documentclass[11pt]{article}

\usepackage{epsfig}
\usepackage{latexsym}
\usepackage{amssymb}

\newcommand{\sect}[1]{\setcounter{equation}{0}\section{#1}}
\renewcommand{\theequation}{\arabic{section}.\arabic{equation}}

\def\be{\begin{equation}}
\def\ee{\end{equation}}
\def\ba{\begin{eqnarray}}
\def\ea{\end{eqnarray}}
\def\dt{\tilde{d}}

\title{{\bf AdS/CFT in the Infinite Momentum Frame}}
\author{{\bf D. Brecher}\thanks{email: Dominic.Brecher@durham.ac.uk}
\\ Centre for Particle Theory, Department of Mathematical Sciences, \\
University of Durham, South Road, Durham DH1 3LE, 
United Kingdom.\\
{\bf A. Chamblin}\thanks{email: chamblin@mit.edu} 
\\ Center for Theoretical Physics, Massachusetts Institute of
Technology, Bldg. 6-304, 
\\ Cambridge, MA 02139, U.S.A.\\
{\bf H.S. Reall}\thanks{email: H.S.Reall@qmw.ac.uk}
 \\ Physics Department, Queen Mary College, \\
 Mile End Road, London E1 4NS, United Kingdom.
\\ \\ Preprint DTP/00/103, QMW-PH/00-15, MIT-CTP-3052}

\date{8 December 2000}

\begin{document}

\maketitle

\begin{abstract}

This paper considers the spacetimes describing pp-waves propagating on
extremal non-dilatonic branes. It is shown that an observer moving along a
geodesic will experience infinite curvature at the horizon of the
brane, which should therefore be regarded as singular.
Taking the decoupling limit of these brane-wave spacetimes 
gives a pp-wave in AdS, the simplest example being the Kaigorodov spacetime. 
It has been conjectured that gravity in this spacetime is dual to a
CFT in the infinite momentum frame with constant momentum density. 
If correct, this implies that the CFT must resolve the singularity of
the bulk spacetime. Evidence in favour of this conjecture is presented. 
The unbroken conformal symmetries determine the scalar 2-point function
up to an arbitrary function of one variable. However,
an AdS/CFT calculation shows that this function is constant 
(to leading order in $1/N^2$) and the result is therefore the same 
as when the full conformal symmetry is unbroken. This paper also
discusses a recently discovered Virasoro symmetry of 
metrics describing pp-waves in AdS and naked singularities in the 
Randall-Sundrum scenario.

\end{abstract}

\sect{Introduction}

In the standard picture~\cite{GT}, when one thinks
of the global causal structure of non-dilatonic $p$-branes one 
thinks of vacuum interpolation.  Far from the brane, the fields 
associated with the brane typically fall
off quickly and the spacetime approaches uncompactified, flat space.  
Near the brane,
the geometry is usually that of a wormhole, or Einstein-Rosen bridge.  As
one approaches the brane, one falls down the throat of the wormhole and
approaches some vacuum which is a valid compactification of the
supergravity theory.  
Thus, while it is often impossible to actually locate the 
``brane worldvolume'' in these effective field theory solutions which
describe gravitating branes, it is still possible to discuss the causal
structure of the configurations.  In particular, it is always the case
that different supersymmetric vacuum solutions of the supergravity theory
emerge in different regions of the full spacetime of a $p$-brane.
Furthermore, when the brane worldvolume is flat there are no 
curvature singularities in the near-horizon region and so we may
extend the spacetime through the horizon.

A natural question is: what happens to this causal structure when one
perturbs the brane?  In this paper, we consider a large class of 
perturbations of the near-horizon region of the brane spacetime, 
corresponding to 
plane fronted parallel gravitons moving tangent to the brane directions.
We find that such Weyl curvature fluctuations always lead to the
appearance of {\it pp-curvature singularities} in the near-horizon
region. In other words, some curvature components diverge in an
orthonormal frame that is parallelly propagated along a geodesic that
reaches the horizon. Thus, these perturbations close off the
spacetime, so that there is no analytic extension of the manifold
through the near-horizon region. Similar results have been previously
noticed for waves on strings \cite{KMR, HY, Myers}. 
It has also been shown that if one
attempts to extend the metric beyond the singular string horizon, then a test
string falling through the horizon would experience a divergent
excitation \cite{Ross}. Therefore, these singularities should be
regarded as a physical boundary to the spacetime.

The pp-wave in the spacetimes that we consider is described by a
harmonic function which can depend on both transverse and worldvolume
brane coordinates. It has been proposed that such solutions can
describe fully localized intersections of a pp-wave with a brane
\cite{Yang}, and explicit solutions for such a localized intersection
have been presented in the near horizon limit \cite{Youm}. We shall
show that it is possible to derive the near horizon solution of
\cite{Youm} simply by requiring that the horizon be free of
pp-singularities. However, we shall then show that the resulting
solution can be related to anti-de Sitter space (AdS) by a coordinate
transformation, so this solution does not really describe a wave.
The only non-trivial non-singular solutions have a  
harmonic function that vanishes on the brane horizon. Such solutions
describe a wave propagating in a direction parallel to the brane, but
at some distance away from the horizon, rather than a wave on the
brane worldvolume. Our results, and the results of \cite{KMR, HY} are
evidence in favour of a generalized ``no-hair'' theorem for time dependent
perturbations of brane horizons.

Taking the decoupling limit \cite{Mal} whilst keeping the momentum 
density of the wave finite, one obtains solutions describing pp-waves
propagating in AdS. The case in which the harmonic function describing
the wave depends only on the transverse direction was discussed in
\cite{cvetic:99}, where it was shown that the Kaigorodov spacetime
\cite{K1,K2} is obtained in the decoupling limit. It was also shown
that if the brane is made non-extremal then the wave can be removed by
a boost. In the limit of extremality, this boost becomes infinite. It
is therefore natural to regard the extremal brane-wave solution
as an infinitely boosted version of the extremal brane solution. In
the decoupling limit, this means that the Kaigorodov spacetime can be
regarded as a infinitely boosted version of AdS. This infinite boost
also induces an infinite boost on the conformal boundary of AdS.
In the light of the AdS/CFT correspondence \cite{Mal}, 
it is therefore natural to conjecture \cite{cvetic:99} that 
string theory in the Kaigorodov spacetime (times a sphere of 
appropriate dimension) is dual to some conformal field theory (CFT) 
in an infinitely boosted frame, i.e., the so-called infinite momentum
frame. Since the momentum density was held fixed in the
decoupling limit, there must be a non-zero background momentum density 
present. If this conjecture is correct then the CFT must resolve the
singularity of the bulk spacetime.

We have calculated the expectation value of the 
CFT energy-momentum tensor that arises from a large
class of solutions describing pp-waves propagating along the
horospheres of AdS. The energy-momentum tensor describes a null
momentum density, with profile fixed by the profile of the bulk
wave. For the Kaigorodov spacetime, the momentum density is indeed
constant, in agreement with the conjecture of \cite{cvetic:99}. 
This background momentum density breaks the conformal
symmetry group of the boundary theory down to some smaller symmetry
group. We show how the isometries of the Kaigorodov spacetime
have a natural interpretation as this subgroup of the conformal
group that leaves the background momentum density invariant.

When conformal invariance is unbroken, it is well-known that the
2-point functions of certain CFT operators are completely
determined. In our case, although some of the conformal invariance is
broken, the dilatation symmetry persists (when combined with a
boost). This highly, but not entirely, constrains the form of the 
2-point functions. For
a spin-0 operator, we show that the symmetry determines the 
2-point function up to an arbitrary function of one variable. 
It is then natural to ask whether we can calculate the 2-point function 
using the prescription of \cite{GKP, witten:98}. This turns out to be 
straightforward, with the exception that the Kaigorodov spacetime is
non-static and therefore cannot be Wick rotated to Euclidean
signature. The AdS/CFT calculation must therefore be carried out on
the Lorentzian spacetime. 

The 2-point function is not uniquely defined in Lorentzian signature. 
However, the arbitrariness takes a form that is characteristic of the infinite
momentum frame, offering further support in favour of the conjecture
of \cite{cvetic:99}. Moreover, the 2-point function turns out to be
independent of the background momentum density. We argue that this
makes sense in the large $N$ limit. 

Another topic addressed in this paper is the Virasoro symmetry group
of anti-de Sitter space. It is well-known that gravity in spacetimes
asymptotic to AdS$_3$ is equivalent to a two-dimensional CFT \cite{BH}.  
It has recently been pointed out that the space of metrics describing
pp-waves propagating on the horospheres of AdS admits an action of the
Virasoro group in {\it any} dimension~\cite{BCG}. We explain how this symmetry
arises from the action of the Virasoro group on 3-dimensional
submanifolds of these spacetimes.

\sect{Singular brane-waves}

Our starting point is the ansatz for the metric of a non-dilatonic
$p$-brane with a pp-wave traveling
along a worldvolume direction, which we take to be $x$.  
The line element can be written as:
\be
 ds^2 = H(r)^{-2/d} ( 2 dx^+ dx^- + F(x^+,x^i,r) dx^{+2} + dx^i dx^i) +
 H(r)^{2/\dt} (dr^2 + r^2 d\Omega_{\dt+1}^2),
\label{metric}
\ee

\noindent where $d = p+1$, $\dt = D-d-2$, $D$ is the dimension of
spacetime and
\be
\label{eqn:dtdef}
\frac{2}{d} + \frac{2}{\dt} = 1,
\ee

\noindent for the branes of interest.  $x^i, ~i=1,\ldots,d-2$, denote the
directions tangent to the brane and transverse to the wave, and we have
written
\be
x^{\pm} = \frac{(x \pm t)}{\sqrt{2}}.
\ee

\noindent $d\Omega_{\dt+1}^2$ is the line-element of a unit
$(\dt+1)$-sphere (or, more generally, any positive Einstein space with
appropriately normalized curvature).  
Where necessary, we will take $m,n = 1,\ldots,\dt+1$
to denote directions in this space.  The function $F(x^+,x^i,r)$
depends on both worldvolume and transverse coordinates so the
wave is potentially fully localized on both the brane and in the transverse
space.

The curvature tensors for this metric are calculated in the
appendix.  For the M2-brane, the four form field strength is
\be
 F_{[4]} = d(H^{-1}) \wedge dx^+ \wedge dx^- \wedge dy,
\label{4-form}
\ee

\noindent where we have written $x^1 = y$.  For the M5-brane, the dual
of the four form field strength is
\be
 *F_{[4]} = d(H^{-1}) \wedge dx^+ \wedge dx^- \wedge dx^1 \wedge dx^2
 \wedge dx^3 \wedge dx^4,
\ee

\noindent and for the D3-brane, the self-dual five form field strength is
\be
 F_{[5]} = G_{[5]} + *G_{[5]},
\ee

\noindent where
\be
 G_{[5]} = d(H^{-1}) \wedge dx^+ \wedge dx^- \wedge dx^1 \wedge dx^2.
\ee

\noindent It is straightforward to verify that these expressions solve the
Einstein equations provided that the function $H(r)$ is harmonic on
the transverse space:
\be
 H(r) = 1 + \frac{k}{r^{\dt}},
\ee

\noindent and the function $F(x^+,x^i,r)$ obeys (e.g.,~\cite{RT,Yang,Youm})
\be
 \frac{\partial^2 F}{\partial r^2} + \frac{(\dt+1)}{r} \frac{\partial
 F}{\partial r} + H(r) \nabla_{\bot}^2 F = 0,
\label{F_eqn}
\ee

\noindent where $\nabla_{\bot}^2$ denotes the Laplacian with respect
to the flat coordinates $x^i$.

There are different types of solution to this equation, but we will not
consider the specific forms here.  Suffice it to say that the usual brane-wave
intersections~\cite{RT} have $F = F(x^+,r)$, the wave being localized in
the transverse space only.  On the other hand, a Ricci-flat
brane~\cite{BP,Bert,FF} can be
constructed by taking $F = F(x^+, x^i)$, in which case the wave is now
localized on the brane worldvolume, but not in the transverse space.
In the near-horizon limit, one obtains solutions describing pp-waves
propagating in anti-de Sitter space. These were discussed in four
dimensions in \cite{Pod}. They have also been used to give a
non-linear discussion of the Randall-Sundrum \cite{RS} model, with the
solution $F = H_{ij}(x^+) x^i x^j$, (where $H_{ij}$ is symmetric and 
trace--free) identified as the massless graviton \cite{CG}.

There is a particular solution which will prove of interest in what follows.  
In the core of the brane, as $r \rightarrow 0$, (\ref{F_eqn}) becomes
\be
\frac{\partial^2 F}{\partial r^2} + \frac{(\dt+1)}{r} \frac{\partial
 F}{\partial r} + \frac{k}{r^{\dt}} \nabla_{\bot}^2 F = 0,
\label{near-horizon}
\ee

\noindent which is solved by~\cite{Youm,BL}
\be
F(x^+,x^i,r) = F_0 (x^+) + F_1 (x^+) \left[ |x|^2 + k \left
( \frac{d-2}{\dt-2} \right) \frac{1}{r^{\dt-2}} \right].
\label{F_exact}
\ee

This solution can, in fact, be extended beyond the near-horizon
region.  By writing $F = F_0(x^+) + F_1 (x^+) \left( |x|^2 + f(r)
\right)$, we find that
\be
F(x^+,x^i,r) = F_0 (x^+) + F_1 (x^+) \left[ |x|^2 - \left(
\frac{d-2}{\dt +2} \right) r^2 + k \left
( \frac{d-2}{\dt-2} \right) \frac{1}{r^{\dt-2}} \right],
\label{F_exact2}
\ee

\noindent is a solution of (\ref{F_eqn}), valid for all $r$. However,
the amplitude of the wave grows as one moves away from the brane, and
destroys the asymptotic flatness of the spacetime. It is therefore
unlikely that this solution is of physical relevance.

\subsection{Near-horizon geometry}

In the near horizon limit of the metric (\ref{metric}), the $1$ in the
harmonic function $H$ can be dropped. Defining a new coordinate by
\be
 z = \frac{d}{\dt} k^{1/2} r^{1-\dt/2},
\ee
then brings the metric to the form
\be
\label{ads}
 ds^2 = \frac{l^2}{z^2} \left(2 dx^+ dx^- + F(x^+,x^i,z) {dx^+}^2 +
dx^i dx^i + dz^2 \right) + (\dt/d)^2 l^2 d\Omega_{\dt+1}^2, 
\ee
where
\be
 l = (d/\dt) k^{1/\dt}.
\ee
In the new coordinate, the equation satisfied by $F$ is
\be
z^{d-1} \frac{\partial}{\partial z} \left( \frac{1}{z^{d-1}} \frac{\partial
F}{\partial z} \right) + \nabla_{\bot}^2 F = 0.
\label{F_ads}
\ee

\noindent Throwing away the $d\Omega_{\dt+1}^2$ piece of the 
spacetime (\ref{ads}),
and with the equation (\ref{F_ads}) governing the function $F$, 
we recognise the metric describing a pp-wave in AdS$_{d+1}$.  
The AdS horizon is now at $z=\infty$, and the conformal
boundary at $z = 0$.  The latter is, in general, 
no longer Minkowski space; and we will have more to say on this below.

\subsection{Supersymmetry considerations}

It is easy enough to see that our brane-waves preserve $1/4$ of the
spacetime supersymmetries, as should be the case for a two-element
intersection.  In the AdS limit considered above, however, one might
expect a supersymmetry enhancement. It is perhaps
surprising that this does not occur. Indeed, the Kaigorodov spacetime
preserves $1/4$ of the supersymmetries~\cite{cvetic:99} and, as we
will now show, this is true of {\it any} pp-wave in AdS. 

Consider, then, the spacetime (\ref{ads}) without the
$d\Omega_{\dt+1}^2$ piece.  The Killing spinor equations for a
$(d+1)$-dimensional supergravity theory with
cosmological constant $\Lambda = - (1/2l^2)d(d-1)$ have the form
\be
\delta \psi_M \equiv {\cal D}_M \epsilon = D_M \epsilon - \frac{1}{2l} \Gamma_M
\epsilon = 0,
\label{eqn:killing}
\ee

\noindent where $D_M$ is the
covariant derivative acting on an arbitrary spinor $\epsilon$.  It is useful to consider the
integrability condition for the existence of Killing spinors, which is
\be
H_{MN} \epsilon \equiv [ {\cal D}_M, {\cal D}_N ] \epsilon =
\frac{1}{4} R_{MNPQ} \Gamma^{PQ} \epsilon + \frac{1}{2l^2} \Gamma_{MN}
\epsilon = 0.
\label{eqn:integrability}
\ee

\noindent Substituting for the Riemann tensor, we find the non-zero
(tangent space) components of this equation to be
\be
H_{+i}\epsilon = - \frac{1}{2}\frac{z^2}{l^2} \left[ \left( \frac{\partial^2 F}{\partial x^i
 \partial x^j} - \frac{1}{z} \frac{\partial F}{\partial z} \delta_{ij}
\right) \Gamma_j - \frac{\partial^2 F}{\partial z
 \partial x^i} \Gamma_z \right] \Gamma_- \epsilon = 0,
\label{int1}
\ee
\be
H_{+z}\epsilon = - \frac{1}{4}\frac{z^2}{l^2} \left[ \left(
\frac{\partial^2 F}{\partial z^2} - \frac{1}{z}
\frac{\partial F}{\partial z} \right) \Gamma_z - \frac{\partial^2 F}{\partial z
 \partial x^i} \Gamma_i \right] \Gamma_- \epsilon = 0,
\label{int2}
\ee

\noindent where all gamma matrices are defined with respect to the
tangent space metric.

The conditions (\ref{int1},\ref{int2}) can be satisfied in one of two ways.  The first is to take
\be
\frac{\partial^2 F}{\partial z \partial x^i} = \frac{\partial^2
F}{\partial z^2} - \frac{1}{z} \frac{\partial F}{\partial z} = \frac{\partial^2 F}{\partial x^i
 \partial x^j} - \frac{1}{z} \frac{\partial F}{\partial z} = 0,
\ee

\noindent which is solved by
\be
F(x^+,x^i,z) = F_0(x^+) + F_1(x^+) \left[ |x|^2 + z^2 \right].
\ee

\noindent This is just the solution (\ref{F_exact}) in the $z$
coordinate.  However, as we shall show below, it in fact corresponds
to pure AdS in an unusual coordinate system, for which the
integrability equations are satisfied identically; for {\it bona fide}
waves, the function $F$ cannot have this form.  So we are led to the
second way of satisfying the integrability conditions
(\ref{int1},\ref{int2}), which is of course to demand that
\be
\Gamma_- \epsilon = 0.
\label{eqn:chirality}
\ee

\noindent This is just the usual chirality condition for a pp-wave to be
supersymmetric, and removes $1/2$ of the supersymmetries.  However,
since this is a necessary, but
not sufficient, condition to ensure the existence of Killing spinors, we must
further consider the equations (\ref{eqn:killing}) explicitly.

The Killing spinor equations
(\ref{eqn:killing}) reduce to the following conditions on $\epsilon$:
\be
 \partial_+ \epsilon = \frac{1}{2z} \Gamma_+ ( 1 + \Gamma_z) \epsilon
+ \frac{1}{4} \left[ \frac{\partial F}{\partial x^i}
\Gamma_i + \left( \frac{\partial F}{\partial z} - \frac{F}{z} \right)
\Gamma_z \right] \Gamma_- \epsilon,
\ee
\be
 \partial_- \epsilon = \frac{1}{4z} (1 - \Gamma_z) \Gamma_- \epsilon,
\ee
\be
 \partial_i \epsilon = \frac{1}{2z} \Gamma_i (1 + \Gamma_z) \epsilon,
\ee
\be
 \partial_z \epsilon = \frac{1}{2z} \Gamma_z \epsilon.
\ee

\noindent Solutions of these equations must satisfy the chirality
condition (\ref{eqn:chirality}).  This removes all trace of the function
$F$, leaving the pure AdS equations, 
for which the Killing spinors were constructed in horospherical
coordinates 
in~\cite{LPT}.  In this case, there are two distinct types of
solution: one a function of all the coordinates, and one
a function of $z$ alone.  
Only the latter type of solution satisfies equation (\ref{eqn:chirality}),
however.  In other words, the sole solution to the Killing spinor equations is
\be
\epsilon = \sqrt{\frac{l}{z}} \epsilon_0,
\label{eqn:spinor}
\ee

\noindent where $\epsilon_0$ is a constant spinor satisfying
\be
(1 + \Gamma_z ) \epsilon_0 = 0,
\ee
\be
\qquad \Gamma_- \epsilon_0 =
0,
\ee

\noindent The pp-wave in AdS thus preserves
only $1/4$ of the supersymmetries.  
It is important to note that this is true regardless of
which coordinates $F$ is a function of.  In particular, it is true of
the Kaigorodov spacetime, to be considered below, as shown
in~\cite{cvetic:99}.  At any rate, the pp-wave in AdS$_5$ should break
the supersymmetry of the dual CFT from ${\cal N}=4$ to ${\cal
N}=1$. This can occur in two different ways: either the pp-wave
changes the background metric of the boundary theory, which explicitly breaks
some of the supersymmetry, or the pp-wave preserves the flat metric of
the boundary theory but corresponds to a quantum state of the boundary
theory in which some of the supersymmetry is spontaneously broken. In
the case of the Kaigorodov metric, the latter possibility occurs, with
the symmetry breaking arising from a non-zero expectation value for
the energy-momentum tensor.

\subsection{Global structure}

Since the pp-wave generates Weyl curvature alone, the only
potentially divergent curvature invariant of our solutions is the
square of the Riemann tensor; and it is easy to see from the curvature
components given in the appendix that this is everywhere
finite. However, it was shown in \cite{Pod} that pp-waves in AdS$_4$
generically 
give rise to pp-curvature singularities at the AdS horizon, i.e., in
an orthonormal frame parallelly propagated along a causal geodesic,
some curvature components will diverge at the AdS horizon.
It is therefore important that we examine the behaviour of the tidal forces
experienced by observers who move along geodesics in brane-wave
spacetimes. 

Consider, then, the geodesic
equations for the spacetime (\ref{metric}).  For simplicity, we take
$d\Omega_{\dt+1}^2$ to be the line-element on the unit
$(\dt+1)$--dimensional sphere, set all the
polar angles to $\pi/2$, and take $\phi$ to be the azimuthal angle.
We further take $t^M$ to denote the unit tangent to a
timelike geodesic, with affine parameter $\lambda$.  Then the Killing
vectors $k = \frac{\partial}{\partial x^-}$ and $m =
\frac{\partial}{\partial \phi}$ generate the
conserved quantities $E = k \cdot t$, and $L = m \cdot t$ respectively.
These give the geodesic equations
\ba
\frac{dx^+}{d\lambda} &=& E H^{2/d}, \\
\frac{d\phi}{d\lambda} &=& \frac{L}{r^2 H^{2/\dt}}.
\ea

\noindent The $x^i$ equation is
\be
\frac{d}{d\lambda} \left( \frac{1}{H^{2/d}} \frac{dx^i}{d\lambda}
\right) = \frac{E^2}{2} H^{2/d} \frac{\partial F}{\partial x^i},
\ee

\noindent and that governing the radial motion is
\[
\frac{d}{d\lambda} \left( H^{2/\dt} \frac{dr}{d\lambda} \right) =
 \frac{1}{2} \frac{H'}{H} \left( \frac{2}{d} +
H^{2/\dt} \left( \frac{dr}{d\lambda} \right)^2 \right) +
\frac{1}{2} E^2 H^{2/d} \frac{\partial F}{\partial r}
\]
\be
+ \frac{1}{2}
\frac{L^2}{r^2 H^{2/\dt}} \left( \frac{2}{r} + \frac{H'}{H} \right),
\label{radial}
\ee

\noindent For
the special case in which $F = F(x^+,x^i)$ and $L=0$, this equation
can be solved to give
\be
\frac{dr}{d\lambda} = \frac{1}{H^{(d-2)/2d}} \sqrt{ c H^{2/d} - 1},
\label{eqn:specialsoln}
\ee

\noindent for some constant $c$.  In general, however, we are unable to
solve (\ref{radial}) due to the $r$-dependence of the function $F$.
Using the timelike condition, the final equation for $x^-$ is
\be
\frac{dx^-}{d\lambda} = -\frac{1}{2E} \left( 1 + \frac{1}{H^{2/d}}
\left( \frac{dx^i}{d\lambda} \right)^2 + H^{2/\dt}
\left( \frac{dr}{d\lambda} \right)^2 + E^2 H^{2/d} F + \frac{L^2}{r^2
H^{2/\dt}} \right).
\ee

As we have already mentioned, we cannot solve this set of equations in
general.  But this does not prevent us from analyzing the
near-horizon geometry of the brane.  To this end, we set
$L=0$ and write the tangent vector to the geodesic as
\be
t^M = \left( E H^{2/d}, \dot{x}^-, \dot{{\bf x}}, \dot{r}, 0, \ldots,0\right),
\ee

\noindent where $\dot{x}^+ = E H^{2/d}$, a dot denotes differentiation
with respect to $\lambda$, ${\bf x} = \{x^i\}$ and we have written the
components in the order $\left( t^+, t^-, t^i, t^r, t^{\theta_1},
\ldots, t^{\theta_{\dt}}, t^{\phi} \right)$. We want to analyze the
behaviour of the components of the Riemann tensor in an orthonormal
frame parallelly propagated along a geodesic with tangent $t^M$. We
must first construct the other vectors of this orthonormal frame.
There are $d-2$ unit normals of the form
\be
n_{i}^M = H^{1/d} \left( 0, - \frac{\dot{x}^i}{E H^{2/d}} , \hat{n}_i, 0,
\ldots, 0 \right),
\ee

\noindent where $\hat{n}_i^j = \delta^j_i$.  The normals satisfy $t
\cdot n_i = 0$, $n_i \cdot n_j = \delta_{ij}$ and it is
easy to check that they are parallelly propagated with respect to $t$:
\be
t \cdot \nabla n_i^M = 0.
\ee

\noindent Although they will prove to be irrelevant to our discussion,
the $\dt+1$ angular normals, $n_m$, are easy to determine.  They
have a single non--zero component
\be
n_m^n = \frac{1}{H^{1/\dt}~ r}~ \frac{1}{\sin \theta_1 \ldots
\sin \theta_{\dt}}~ \delta^n_m.
\ee

\noindent The remaining two normals of our parallelly propagated frame
are somewhat harder to determine.  Denoting them by $p$ and $q$, we have
\be
 p^M = \sin \mu (\lambda) \left( EH^{2/d}, \dot{x}^- + \frac{1}{E},
\dot{{\bf x}} , \dot{r} ,0, \ldots ,0 \right) + \cos \mu (\lambda)
\left(0, \frac{\dot{r}}{E} H^{1/\dt}, {\bf 0}, -H^{-1/\dt},0, \ldots
,0 \right),
\ee
\be
 q^M = \cos \mu (\lambda) \left( EH^{2/d}, \dot{x}^- + \frac{1}{E},
\dot{{\bf x}} , \dot{r} ,0, \ldots ,0 \right) - \sin \mu (\lambda)
\left(0, \frac{\dot{r}}{E} H^{1/\dt}, {\bf 0}, -H^{-1/\dt},0, \ldots
,0 \right), 
\ee

\noindent where $\mu (\lambda)$ obeys
\be
 \frac{d\mu}{d\lambda} = -\frac{H'}{d H} H^{-1/\dt},
\ee

\noindent which gives $\mu = {\rm constant}$ far from the brane ($r \rightarrow
\infty$), and $\mu = \lambda/l + {\rm constant}$ 
in the near horizon region ($r \rightarrow 0$). These normals are
slight generalizations of those given by Podolsk\'{y}~\cite{Pod} in a
study of pp-waves in $\mbox{AdS}_4$.

\subsection{Waves in AdS$_{d+1}$}

Before analyzing the global structure of our brane-waves, it will be
instructive to consider the simpler case of waves in AdS$_{d+1}
\times S^{\dt+1}$, the near-horizon geometry of the metric
(\ref{metric}). Note that it is not sufficient to just consider waves
in AdS when studying pp-singularities at the horizon of the brane-wave
spacetime. This is because in taking the near horizon limit, one has
to take the limit of the components of the curvature in a parallelly
propagated frame, e.g., of objects like $t^M p^N t^ P p^Q
R_{MNPQ}$. However, in the AdS limit, one takes the limits of $t^M$,
$p^N$ and $R_{MNPQ}$ separately and then multiplies them. As we shall
see, the product of these limits is not necessarily equal to the limit of the
product. Therefore it is necessary to examine the brane-wave
spacetime and the pp-wave in AdS separately.

The presence of the sphere is irrelevant, so 
we will essentially be considering waves in AdS$_{d+1}$, and the
results are again a slight generalization of those
in~\cite{Pod}. To look at the AdS case, as opposed to our brane-wave
spacetime (\ref{metric}), we can simply set $H(r) = k/r^{\dt}$ in the
expressions for the curvature components, tangent vector and
normals. In this case the equation
of motion (\ref{near-horizon}) valid in the core of the brane, is
exact.  Analysis of the curvature
components in our parallelly propagated frame is then fairly straightforward.

There are a number of potential pp-singularities, 
but they are all encoded in the quantities\footnote{
The reader may be worried that these expressions appear to depend on
$\dt$, which is related to the dimension of the sphere. However,
equation (\ref{eqn:dtdef}) can be used to eliminate $\dt$.}
\ba
{\cal A}_+ &=& \frac{E^2}{2} \left( \frac{k}{r^{\dt}}
\right)^{1-4/\dt} \left[ F'' + (\dt-1) \frac{F'}{r} \right], \\
{\cal A}_i &=& \frac{E^2}{2} \left( \frac{k}{r^{\dt}}
\right)^{3/2-4/\dt} \frac{\partial^2 F}{\partial r \partial x^i}, \\
{\cal M}_{ij} &=& \frac{E^2}{2} \left( \frac{k}{r^{\dt}} \right)^{2-4/\dt}
\left[ \frac{\partial^2 F}{\partial x^i \partial x^j} + \delta_{ij}
\frac{\dt}{d} \frac{r^{\dt}}{k} \frac{F'}{r} \right],
\ea

\noindent which are potentially divergent near the AdS horizon
($r \rightarrow 0$).  \emph{All} components
of the curvature tensor in our parallelly propagated frame depend on
these three quantities alone.  For example,
\be
R_{(t)(p)(t)(p)} \equiv R_{MNPQ} t^M p^N t^P p^Q =
\frac{1}{l^2} - {\cal A}_+ \cos^2 (\lambda/l),
\label{curvature1}
\ee
\be
R_{(t)(p)(t)(i)} \equiv R_{MNPQ} t^M p^N t^P n^Q_i 
= {\cal A}_i \cos (\lambda/l),
\label{curvature2}
\ee
\be
R_{(t)(i)(t)(j)} \equiv R_{MNPQ} t^M n^N_i t^P n^Q_j =
\frac{\delta_{ij}}{l^2} - {\cal M}_{ij}.
\label{curvature3}
\ee

\noindent For the waves in AdS under consideration here, these
expressions are exact.

The question is, how do the specific choices of the function $F$
affect the existence, or otherwise, of pp-singularities? It is easy
to see that the partially localized solutions with $F=F(x^+,r)$ or
$F=F(x^+,x^i)$ are indeed singular\footnote{The special case
$F(x^+,x^i) = a_i(x^+) x^i + F_0(x^+)$ will be discussed below. Note
that the case $F=F(x^+,x^i)$ includes the solution
$F=H_{ij}(x^+) x^i x^j$, describing the Randall-Sundrum \cite{RS} zero mode, 
which was previously thought to be
non-singular~\cite{CG}.  In this case, we have ${\cal M}_{ij} \sim
H_{ij}(x^+)/r^{2\dt-4}$.  Of course, in AdS$_3$ this term
vanishes since, in this case, $H_{ij} \rightarrow H_{11} = 0$ by
virtue of the tracelessness of $H_{ij}$.  However, we do not really
have a pp-wave in this case, since there can be no propagating
gravitational degrees of freedom in three dimensions.  In
higher dimensions, the zero mode does, in fact, generate
singularities at the AdS horizon.} as $r\rightarrow 0$.
However, one might suspect that there exist 
\emph{fully} localized solutions
with $F = F(x^+,x^i,r)$  that are entirely 
non-singular.  We will see that this is not true.

Absence of pp-singularities requires that all curvature components
remain finite as $r \rightarrow 0$. From equation
(\ref{curvature2}) and the expression for ${\cal A}_i$, 
this implies that $ \partial_r F = g_0(r,x^+) +
{\cal O}(r^{3\dt/2-4})$ as $r \rightarrow 0$, 
where we are using the notation $f={\cal
O}(g)$ to mean that there exists some function $h(x^+,x^i)$ such that
$f/g \le h$ as $r \rightarrow 0$. 
Integrating gives $F= g_1 (r,x^+) + g_2 (x^i,x^+) + {\cal
O}(r^{3\dt/2-3})$, for some $g_1$ and $g_2$. Substituting this into
the equation of motion for $F$ gives
\be
 \frac{1}{r} \frac{\partial}{\partial r} \left( r^{\dt +1}
\frac{\partial g_1}{\partial r} \right) + k \nabla_{\bot}^2 g_2 =
{\cal O}(r^{5 \dt/2-5}),
\ee
from which it follows that 
\be 
\label{eqn:del2g2}
 \nabla_{\bot}^2 g_2 = \frac{2}{k} (\dt-2)  F_1(x^+)
\ee
for some function $F_1$. Integrating with respect to $r$ gives
\be
\label{eqn:Fintermed}
 F(x^+,x^i,r) = F_0(x^+) + \frac{F_1 (x^+)}{r^{\dt-2}} + 
 \frac{F_2(x^+)}{r^{\dt}} + g_2(x^+,x^i) + {\cal O}(r^{3\dt/2-3}).
\ee
Substituting this into equation (\ref{curvature1}) then yields $F_2
\equiv 0$ in order for the curvature to remain finite as $r
\rightarrow 0$.
Finally, substituting into equation (\ref{curvature3}) and requiring 
finiteness as $r \rightarrow 0$ gives
\be
\label{eqn:g2eq}
 \frac{\partial^2 g_2}{\partial x^i \partial x^j} =
\frac{\dt(\dt-2)}{kd} F_1(x_+) \delta_{ij},
\ee
which is consistent with equation (\ref{eqn:del2g2}). Solving this
equation then gives the final result for $F$:
\be
\label{eqn:Fsol} 
F(x^+,x^i,r) = F_0(x^+) + F_1(x^+) \left[ |x|^2 + k \left
 ( \frac{d-2}{\dt-2} \right) \frac{1}{r^{\dt-2}} \right] + a_i(x^+) x^i
 + \Delta(x^+,x^i,r),
\ee

\noindent where we have rescaled $F_1$. $\Delta$ denotes a possible
correction term of order $r^{3\dt/2-3}$. Note that $F_0$ can be gauged
away by redefining $x^-$. The $a_i(x^+) x^i$ term can be also be
gauged away, as we shall explain below. So, for the time being, we shall
set $F_0 = a_i = 0$.
We should emphasize that equation (\ref{eqn:Fsol}) is the only solution 
free of pp-singularities.
If $\Delta$ is identically zero then 
this is just the near-horizon solution (\ref{F_exact}) discussed in
\cite{Youm,BL} (with an additional term linear in $x^i$),
i.e., it is an exact solution of the equation of
motion (\ref{near-horizon}) for pp-waves in AdS. 
It follows that the correction $\Delta$ must
also be an exact solution of equation (\ref{near-horizon}), but this
correction is negligible near the horizon. If one had just this
correction term, then one would also have a solution with no
pp-singularities at the horizon. This is easily understood: a solution
of ${\cal O}(r^{3\dt/2-3})$ vanishes at the horizon sufficiently fast
that the metric is just like pure AdS as far as geodesics and
curvature tensors are concerned\footnote{
This is not strictly true in general since second derivatives of  
$\Delta$ would naively be expected to behave as $r^{3\dt/2-5}$, which
vanishes at $r=0$ only if $\dt>3$, i.e., $d<6$. However, it is
probably possible to derive stronger bounds on $\Delta$.}. 
In other words, such a solution 
would represent some perturbation of pure AdS that does not extend as 
far as the AdS horizon. It seems unlikely that such a solution would
correspond to the near-horizon geometry of an intersection of a brane
and a pp-wave. The only candidate for such a solution is therefore
the solution (\ref{F_exact}). However, it is easy to use the curvature
tensors in the appendix to see that this solution is Weyl flat,
which implies that it is locally isometric to pure AdS and therefore
does not describe a pp-wave in AdS. We shall present the coordinate
transformation that gauges away this solution later. Given that this
solution is pure gauge, it seems rather doubtful that it describes
the near-horizon geometry of a fully localized intersection of a brane
and pp-wave, as suggested in \cite{Youm}.

Now, as promised, let us re-instate the term $a_i(x^+) x^i$. The
coordinate transformations that removes the first and second terms in
\ref{eqn:Fsol} do not affect the form of the third term ($a_i$ is
just multiplied by a function of $x^+$), and the change in $F$ is the
only change in the metric. So once we have removed
the first two terms in $F$, 
we can do the following coordinate transformation:
\be
 x^- \rightarrow x^- + b(x^+) + c_i(x^+) x^i, \qquad x^i \rightarrow
 x^i + d^i (x^+).
\ee
It is easy to see that $b(x^+),c_i(x^+)$ and $d_i(x^+)$ can be chosen
to cancel the $a_i(x^+) x^i$ term in $F$.

To summarize, we have shown that there are no non-singular
pp-waves in AdS$_{d+1}$ with metric of the form (\ref{ads}), 
except possibly solutions that vanish sufficiently fast near the AdS
horizon.

\subsection{Brane-waves}

We will now generalize the above analysis to the more complicated case
of our brane-waves, described by the metric (\ref{metric}).
Concentrating on the same curvature components as in
(\ref{curvature1}), (\ref{curvature2}) and (\ref{curvature3}) above,
in the near-horizon limit, $r\rightarrow 0$, we find
\be
R_{(t)(p)(t)(p)} \rightarrow \frac{1}{l^2} - {\cal A}_+ \cos^2 (\lambda/l) +
\frac{\dt}{d} (1+\dt) \frac{r^{\dt}}{k} \left( \frac{\dot{r}}{r}
\right)^2 \cos^2 (\lambda/l),
\label{branecurvature1}
\ee
\be
R_{(t)(p)(t)(i)} \rightarrow {\cal A}_i \cos (\lambda/l),
\label{branecurvature2}
\ee
\be
R_{(t)(i)(t)(j)} \rightarrow \frac{\delta_{ij}}{l^2} - {\cal M}_{ij} +
\delta_{ij} \frac{\dt}{d} (1+\dt) \frac{r^{\dt}}{k} \left( \frac{\dot{r}}{r}
\right)^2.
\label{branecurvature3}
\ee

\noindent Note that there are now extra terms involving $\dot{r}$. It
possible that these terms could cancel the divergences coming from the
other terms. This emphasizes the importance of examining the full
brane-wave spacetime, and not just the near-horizon geometry.
We have retained only the leading order divergences, and
assumed that $\dot{r}$ diverges as $r \rightarrow 0$. (If $\dot{r}$
remains finite then the extra contributions above are clearly finite
and cancellations will not occur.)

Using equation (\ref{branecurvature2}), we can obtain equation
(\ref{eqn:Fintermed}) as above. However, there will now be an extra
${\cal O}(r^2)$  correction term on the right hand side, arising from
neglecting the $1$ in $H$ in deriving equation
(\ref{near-horizon}). This is subleading order for the M5-brane but leading
order for the other branes\footnote{Once again, it may be possible to
derive better bounds on the correction term.}.

It is now possible that the term in $F$ of order $r^{-\dt}$ may be
consistent with finite curvature provided that the $\dot{r}$ terms
cancel the divergences arising from this term. Equation
(\ref{branecurvature1}) shows that $\dot{r}$ would have to be of
order $r^{2-3\dt/2}$ as $r\rightarrow 0$ in order to cancel the
divergence in ${\cal A}_+$. To see whether this is possible, we need
to examine the geodesic equation (\ref{radial}). The dominant term on
the right hand side would be the second term (involving
$\dot{r}^2$). Neglecting the other terms, and integrating then gives
$\dot{r} \sim r^{2-\dt/2}$ (this could also be obtained from equation
(\ref{eqn:specialsoln})), which is a contradiction. It is therefore not
possible for the $\dot{r}$ terms to cancel the divergence arising from
the $r^{-\dt}$ term in $F$, so $F_2$ has to be set to zero in
order to obtain finite curvature, just as above.

Now consider equation (\ref{branecurvature3}). 
An identical argument shows that the $\dot{r}$ term cannot
cancel the divergence coming from ${\cal M}_{ij}$, so one obtains
equation (\ref{eqn:g2eq}) as above.  Putting everything together, one
obtains the same solution (\ref{eqn:Fsol}) but with the exception that
the correction term is ${\cal O}(r^2)$ for the M2 and D3 branes. (Note
that such a term occurs in equation (\ref{F_exact2}).)

Thus, we conclude that the only non-singular brane-waves must have $F$ of
the form (\ref{F_exact}) near the brane.  Since this form for $F$ can be
gauged away, we conclude that \emph{there are
no non-singular pp-waves on the non-dilatonic branes},
of the supergravity theories that we have been considering.
This is reminiscent of the results concerning waves on strings in five and six
dimensions~\cite{KMR,HY,Myers}, 
in which similar singularities in the curvature
components at the horizon of the string were found.

\sect{Virasoro symmetry}

Before moving on to the cases of interest to us, we will review some
relevant points concerning AdS$_3$ and its two--dimensional CFT dual.

\subsection{AdS$_3$, Virasoro symmetry and $(2+1)$-dimensional ``pp-waves''}

Consider AdS$_3$ in horospherical coordinates\footnote{
We have rescaled $x^{\pm}$ by a factor of $\sqrt{2}$.}
\be
ds^2 = \frac{l^2}{z^2} \left( dx^+ dx^- + dz^2 \right).
\label{ads_3}
\ee

\noindent Since AdS$_3$ is the group manifold $SL(2,\mathbb{R})$, the metric (\ref{ads_3}) has the
isometry group $SL(2,\mathbb{R})_L \otimes SL(2,\mathbb{R})_R = SO(2,2)$.  The
$SL(2,\mathbb{R})_L$ action is given by:
\ba
\label{sl2r_u}
x^+ &\rightarrow& f(x^+), \nonumber\\
x^- &\rightarrow& x^- - \frac{1}{2} z^2 \frac{f''}{f'}, \\
z &\rightarrow& z \sqrt{f'}, \nonumber
\ea

\noindent where
\be
f(x^+) = \frac{ax^+ + b}{cx^+ + d} \in SL(2,\mathbb{R}),
\ee

\noindent and a prime denotes a derivative with respect to $x^+$.  Note
that elements of $SL(2,\mathbb{R})$ have a vanishing Schwarzian derivative:
\be
\label{eqn:schwarzian}
\{ f,x^+\} = \frac{f'''}{f'} - \frac{3}{2} \left( \frac{f''}{f'} \right)^2 = 0.
\ee

\noindent The $SL(2,\mathbb{R})_R$ action has the same form as (\ref{sl2r_u}),
but with $x^+$ and $x^-$ interchanged:
\ba
x^- &\rightarrow& f(x^-), \nonumber\\
x^+ &\rightarrow& x^+ - \frac{1}{2} z^2 \frac{f''}{f'}, \\
z &\rightarrow& z \sqrt{f'}, \nonumber
\label{sl2r_v}
\ea

\noindent where now
\be
f(x^-) = \frac{ax^- + b}{cx^- + d} \in SL(2,\mathbb{R}),
\ee

\noindent and a prime denotes a derivative with respect to $x^-$.
This isometry group acts on the boundary as the global conformal
group of two-dimensional Minkowski space, generated by two copies of
the Virasoro generators $L_0$ and $L_{\pm 1}$.

Now the AdS/CFT correspondence asserts that the AdS isometry group is
identical to the conformal group on the boundary.  In two dimensions,
the latter is (two copies of)
the infinite dimensional Virasoro group, so there is a puzzle here:
how can the finite-dimensional isometry group be equivalent to the 
infinite-dimensional conformal
group on the boundary?  The puzzle was resolved long ago by Brown and
Henneaux~\cite{BH}.  The isometries (\ref{sl2r_u}) and (\ref{sl2r_v}) 
leave the metric
unchanged, and the corresponding generators will annihilate the vacuum
state in the quantum theory.  These isometries must, however, be
supplemented by the infinite number of diffeomorphisms
which \emph{change} the metric, but leave the asymptotic form of AdS$_3$
invariant up to a conformal factor.  An analysis of the (modes of the)
generators of these allowed diffeomorphisms shows that they obey (two
copies of) the Virasoro algebra.  The asymptotic symmetry group of
AdS$_3$ is thus just the conformal group in $1+1$ dimensions, as expected.

The induced metric on the boundary is
\be
ds^2 = \frac{l^2}{z^2} dx^+ dx^-,
\ee

\noindent the conformal transformations of which are $x^+ \rightarrow
f(x^+)$ and $x^- \rightarrow f(x^-)$.  In the bulk, then, the asymptotic
structure-preserving diffeomorphisms are precisely 
those of (\ref{sl2r_u}) and
(\ref{sl2r_v}) but for arbitrary functions $f(x^+)$ and $f(x^-)$.  
By arbitrary here, we mean
$\{f,x^+\}, \{f,x^-\} \ne 0$.  
Under the former transformation, we find
\be
ds^2 \rightarrow \frac{l^2}{z^2} \left( dx^+ dx^- + 
F(x^+) z^2 dx^{+2} + dz^2 \right),
\label{ads3_V}
\ee

\noindent where
\be
F(x^+) = -\frac{1}{2} \{f,x^+\}.
\label{Fu}
\ee

\noindent  As promised, these general
transformations are not isometries of the metric, but do leave the
boundary manifold unchanged up to a conformal factor (equal to $f'$).

There are two points to note about the metric (\ref{ads3_V}).  Firstly, it looks like a pp-wave
in AdS$_3$.  However, there are no propagating degrees of freedom in
$(2+1)$--dimensional gravity, so what we mean by this is the following.
The metric of a wave in AdS$_3$ would be of the form
\be
ds^2 = \frac{l^2}{z^2} \left( dx^+ dx^- + F(x^+,z) dx^{+2} + dz^2 \right),
\label{ads3_wave}
\ee

\noindent which is a solution of Einstein's equations with negative
cosmological constant if and only if
\be
\label{eqn:Feq3d}
\partial_z \left( \frac{1}{z} \partial_z F \right) = 0,
\ee

\noindent implying
\be
F(x^+,z) = F_0(x^+) z^2 + F_1(x^+),
\ee

\noindent for arbitrary functions $F_0$ and $F_1$. Now, $F_1$ can be
gauged away by a redefinition of $x^-$, leaving a metric of the form
(\ref{ads3_V}), which we know is obtained from pure AdS$_3$ by a
coordinate transformation of the form (\ref{sl2r_u}) with arbitrary
$f(x^+)$. Therefore there are no pp-waves in AdS$_3$, as expected.

The metric (\ref{ads3_V}) is locally isometric to AdS$_3$. This is
reminiscent of the BTZ black hole \cite{BTZ1,BTZ2}. In fact, when
$F(x^+)={\rm constant}$, it is straightforward to obtain the extreme 
BTZ black hole from the metric (\ref{ads3_V}) \cite{BL}. 
To see this, we compactify along the direction of
propagation of the wave, taking the period to be $2\pi$.  Calling this
coordinate $\phi$, taking $F(x^+) = 8GMl^2$, and rewriting in terms of
$r = 1/z$, the metric (\ref{ads3_V}) becomes
\be
\frac{ds^2}{l^2} = -(r^2 - 4GMl^2) dt^2 + (r^2 + 4GMl^2) d\phi^2 -
4GMl^2 dt ~d\phi + \frac{dr^2}{r^2}.
\ee

\noindent Rescaling $t \rightarrow l^2 t$ and $\phi \rightarrow l
\phi$ gives the metric of the extreme $Ml = J$ BTZ black hole~\cite{BTZ1,BTZ2}:
\be
ds^2 = -N^2 dt^2 + \rho^2 \left( N^{\phi} dt + d\phi \right)^2 +
\frac{r^2}{N^2 \rho^2} dr^2,
\ee

\noindent where
\[
\rho^2 = r^2 + 4GMl^2,
\]
\[
N^2 = r^4/(l \rho)^2,
\]
\[
N^{\phi} = -4GMl/\rho^2.
\]

\noindent So there is a nice correspondence: by periodically
identifying one of the coordinates, AdS$_3$ in Poincar\'{e}
coordinates becomes the $M=J=0$ BTZ black hole~\cite{BTZ1,BTZ2}; and a ``wave''
in AdS$_3$ becomes the extreme $Ml=J$ BTZ hole~\cite{cvetic:99,BL}.  

\subsection{Virasoro symmetry in higher dimensions}

The isometry group of AdS$_3$ combines with a subgroup of general 
diffeomorphisms to give the Virasoro group. In higher dimensions this
is not possible.  
However, it has recently been shown \cite{BCG} that the space of
metrics of the form
\be
\label{eqn:ppwaveads} 
 ds^2 = \frac{l^2}{z^2} \left( dx^+ dx^- + F(x^+,x^i,z) {dx^+}^2 +
 dx^i dx^i + dz^2 \right)
\ee
does possess an infinite dimensional symmetry group, namely
\ba
\label{eqn:vira}
 x^+ & \rightarrow & f(x^+), \nonumber \\ 
 x^- & \rightarrow & x^- - \frac{1}{2}(|x|^2 + z^2)\frac{f''}{f'}, \nonumber \\
 x^i & \rightarrow & x^i \sqrt{f'}, \\
 z   & \rightarrow & z \sqrt{f'}, \nonumber
\ea
with the function $F$ shifting according to
\be
 F(x^+,x^i,z) \rightarrow F(x^+,x^i,z)f' -\frac{1}{2}(|x|^2 + z^2)
\{f,x^+\},
\ee
with the Schwarzian derivative defined in (\ref{eqn:schwarzian}). If one
starts from pure AdS$_{d+1}$ ($F \equiv 0$) then the change of coordinates
(\ref{eqn:vira}) takes one to a pp-wave metric with
\be
\label{H}
 F(x^+,x^i,z) = F(x^+) (|x|^2 + z^2),
\ee
for some arbitrary $F(x^+)$. Writing this in terms of the coordinate
$r$ and using the relation (\ref{eqn:dtdef}) brings $F$ to the form
(\ref{F_exact}) without the $F_0$ and $a_i x^i$ pieces 
(which can be gauged away anyway). Therefore the solution
(\ref{F_exact}) just describes pure AdS in unusual coordinates, which
justifies our claim that this solution is pure gauge.

It was proved in \cite{BCG} that the transformations (\ref{eqn:vira}) do
indeed generate the Virasoro group. It is natural to ask why this
symmetry exists. To answer this question, we shall need to discuss two
different sets of coordinate on AdS. First, recall that AdS is defined
as a hyperboloid embedded in flat space with two time directions:
\be
 ds^ 2 = du dv + dX^+ dX^- + dX^i dX^i,
\ee
\be
  uv + X^+ X^- + X^i X^i = -l^2.
\ee
If one now defines $x^{\pm} = X^{\pm}/u$ and $x^i = X^i/u$, and
eliminates $v$, then one obtains the metric for AdS in horospherical
coordinates:
\be
\label{eqn:adshoro}
 ds^2 = l^2 \frac{du^2}{u^2} + u^2 (dx^+ dx^- + dx^i dx^i).
\ee
A second set of coordinate on AdS can be constructed based on the
observation that the surfaces $X^i={\rm constant}$ have AdS$_3$
geometry. Define new coordinates by
$X^1 = l \sinh y \, \cos \theta_1$, $X^2 = l \sinh y \, \cos \theta_1
\sin \theta_2$ etc., $u = l \cosh y \, \tilde{u}$, $v= l\cosh y \,
\tilde{v}$, $X^{\pm} = l \cosh y \, \tilde{X}^{\pm}$, which gives
\be
 ds^2 = l^2(dy^2 + \cosh^2 y (d\tilde{u} d\tilde{v} + d\tilde{X}^+
d\tilde{X}^- ) + \sinh^2 y \, d\Omega^2), 
\ee
\be
 \tilde{u} \tilde{v} + \tilde{X}^+ \tilde{X}^- = -1,
\ee
which makes it clear that surfaces of constant $y$ and $\Omega$ are
AdS$_3$ hyperboloids. Now introduce horospherical coordinate on each
of these hyperboloids: $x^{\pm} = \tilde{X}^{\pm}/\tilde{u}$ and
eliminate $\tilde{v}$ to get
\be
\label{eqn:ads3slice}
 ds^2 = l^2 \left( dy^2 + \cosh^2 y
\left(\frac{d\tilde{u}^2}{\tilde{u}^2} + \tilde{u}^2 dx^+ dx^- \right)
+ \sinh^2 y \, d\Omega^2 \right).
\ee
It is straightforward to work out how the coordinate systems used in
(\ref{eqn:adshoro}) and (\ref{eqn:ads3slice}) are related. The coordinates
$x^{\pm}$ are the same for the two metrics. The other coordinates are
related by $u = l\cosh y \tilde{u}$ and $x^1 u = l\sinh y \cos
\theta_1$ etc. 

Consider now a pp-wave spacetime, with metric
\ba
\label{eqn:pphoro}
 ds^2 &=& l^2 \frac{du^2}{u^2} + u^2 \left( dx^+ dx^- + F(x^+,x^i,u)
        {dx^+}^2 + dx^i dx^i \right) \nonumber \\
      &=&  ds^2 (AdS) + u^2 F(x^+,x^i,u) {dx^+}^2.
\ea
This is written in the horospherical coordinates. Now transform to the
other coordinate system:
\ba
\label{eqn:ppnew}
 ds^2 &=& ds^2 (AdS) + l^2 \cosh^2 y \, \tilde{u}^2 \, F {dx^+}^2 \\
      &=& l^2 \left( dy^2 + \cosh^2 y
 \left(\frac{d\tilde{u}^2}{\tilde{u}^2} + \tilde{u}^2 ( dx^+ dx^- +
  F {dx^+}^2 )  \right)
+ \sinh^2 y \, d\Omega^2 \right). \nonumber
\ea 
It now looks like the surfaces of constant $y$ and $\Omega$ have the
geometry of pp-waves in AdS$_3$, which were discussed above. However,
we have to be slightly careful because $F$ depends on the coordinates
$y$ and $\Omega$, so these pp-waves will be off-shell in three
dimensions (i.e., $F$ will not satisfy equation (\ref{eqn:Feq3d})).
Now consider the Virasoro symmetry (\ref{eqn:vira})
acting on the metric (\ref{eqn:pphoro}) (with $u = l/z$). Writing this in
terms of the new coordinates gives
\ba
 x^+ &\rightarrow& f(x^+), \nonumber\\
 x^- &\rightarrow&  x^- - \frac{1}{2\tilde{u}^2} \frac{f''}{f'}, \\
 \tilde{u} &\rightarrow&  \frac{\tilde{u}}{\sqrt{f'}}, \nonumber
\ea
with $y$ and $\Omega$ unchanged and 
\be
 F \rightarrow F f' - \frac{1}{2\tilde{u}^2} \{f,x^+\}.
\ee
This is clearly nothing but the Virasoro symmetry acting on the three
dimensional surfaces of constant $y$ and $\Omega$. Thus the existence
of the higher dimensional Virasoro symmetry arises from the existence
of the Virasoro symmetry for three dimensional asymptotically AdS 
submanifolds\footnote{
Actually these submanifold are only asymptotically AdS when $F
\rightarrow 0$ as $\tilde{u} \rightarrow \infty$, which implies $F
\rightarrow 0$ as $u \rightarrow \infty$, which implies that
the higher dimensional spacetime is also asymptotically AdS. If this
requirement is not satisfied then the action of the Virasoro symmetry
on the three dimensional submanifolds is a generalization of the
usual action to certain spacetimes that are not asymptotic to AdS$_3$.}.

Note that the higher dimensional pp-wave (\ref{eqn:ppwaveads}) can only
be asymptotic to AdS when $F \rightarrow 0$ as $z \rightarrow
0$. More general pp-waves will not have Minkowski space as their
conformal boundary at infinity, and therefore be of less interest from
the point of view of AdS/CFT. A Virasoro
transformation will introduce a term $|x|^2 \{f,x^+\}$ into $F$. This
destroys the preferred coordinate system on the boundary at infinity
(although the boundary is, of course, still flat space since this is
only a coordinate transformation), and therefore does not correspond
to a conformal transformation on the boundary. The only
transformations that preserve the preferred flat coordinates on the
boundary are those with $\{f,x^+\}=0$, which give an $SL(2,R)$
subgroup of the conformal group on the boundary.

\sect{Boundary energy-momentum tensor}

We have shown that a pp-wave propagating on a non-dilatonic brane has
pp-curvature singularities at the horizon.  The near-horizon geometry
of such spacetimes is a pp-wave in AdS space and we will now consider the
AdS/CFT correspondence in such cases. If this makes sense then the CFT
provides a resolution of the singularity. The easiest thing to compute is
the CFT energy-momentum tensor, making use of the boundary
counterterm method of~\cite{witten:98, henningson:98, tseytlin:98,
henningson:00, BK, emparan:99, KLS}. Our starting point is the metric
(\ref{eqn:ppwaveads}), describing a pp-wave in AdS$_{d+1}$.  
The boundary is at $z=0$, and the induced metric on surfaces of
constant $z$ is
\be
\left. ds^2 \right|_{z=\mbox{\footnotesize const.}} = g_{ij}(x)
dx^i dx^j = \frac{l^2}{z^2} \left( dx^+ dx^- + F(x^+,x^i,z)
{dx^+}^2 + dx^i dx^i \right),
\ee

\noindent where $i,j = 0, \ldots, d-1$.  The outward unit normal
to such a surface is $n = -(l/z) dz$. Using the formulae from
\cite{KLS}, the energy-momentum tensor is given by
\be
 <T_{ij}> = \frac{1}{8\pi G_{d+1}} \left[ -\left(K_{ij} - K g_{ij}
\right) - \frac{d-1}{l} g_{ij} + \frac{l}{d-2} \left( R_{ij} -
\frac{1}{2} R g_{ij} \right) + \ldots \right],
\ee
where $R_{ij}$ and $R$ are the Ricci tensor and Ricci scalar
associated with $g_{ij}$ and $K_{ij}$ denotes the tangential
components of the extrinsic curvature of the boundary, defined by
\be
 K_{\mu\nu} = h_\mu^\rho h_\nu^\sigma \nabla_\rho n_\sigma,
\ee
where $h^\mu_\nu = \delta^\mu_\nu - n^\mu n_\nu$. For $d \ge 6$ one
needs further counterterms.

This calculation will yield the CFT energy-momentum tensor in the
conformal frame where the metric is $g_{ij}$, which includes the
prefactor $l^2/z^2$. To get rid of this factor, one must perform a
Weyl transformation to obtain the metric $\bar{g}_{ij} = (z^2/l^2)
g_{ij}$. In the new conformal frame, the energy momentum tensor is 
$\bar{T}_{ij} = (l/z)^{d-2} T_{ij}$ \cite{wald}, and the metric is
\be
 ds^2 = \bar{g}_{ij} dx^i dx^j = 
 dx^+ dx^- + F(x^+,x^i,0) {dx^+}^2 + dx^i dx^i,
\ee
which is manifestly flat if $F(x^+,x^i,0)=0$,  which is what we shall
usually assume. The counterterm calculation is straightforward. The
only non-vanishing component of the CFT energy-momentum tensor turns out
to be
\be
 <\bar{T}_{++}> = \frac{l^{d-1}}{16 \pi G_{d+1}} \lim_{z\rightarrow 0}
 \frac{1}{z^{d-1}} \left[\frac{\partial F}{\partial z} - \frac{z}{d-2}
 \nabla_{\bot}^2 F - \frac{z^3}{(d-2)^2 (d-4)}
 \left(\nabla_{\bot}^2\right)^2 F + \ldots \right],
\ee
where we have included the first three boundary counterterms. However,
not all three are present for low dimensional examples. 
For $d=2$, only the first term above should be kept; for $d=3,4$, the first
two terms are needed, one keeps the first two terms; and for $d=5,6$,
one needs the third term too. Higher order terms would occur in higher
dimensional examples.

Finally, we must write everything in terms of the parameters of the
dual CFT. For $d=2$, the CFT has central charge $c=3l/(2G_3)$
\cite{BH}, so we obtain
\be
 <T_{++}> = \frac{c}{24 \pi} \lim_{z \rightarrow 0} \frac{1}{z} \frac{\partial
 F}{\partial z},
\ee
where we have dropped the bar on $T_{++}$ since we shall always be
working in the same conformal frame. In the examples obtained from
non-dilatonic branes, namely $d=3,4,6$, the following formulae apply
\cite{Mal}
\be
 \frac{l^2}{G_4} = \frac{2 \sqrt{2} N^{3/2}}{3},
\ee
\be
 \frac{l^3}{G_5} = \frac{2 N^2}{\pi},
\ee
\be
 \frac{l^5}{G_7} = \frac{16 N^3}{3\pi^2},
\ee
where $N$ is the number of branes present before the decoupling limit
is taken. Substituting into $<T_{++}>$ then gives
\be
 <T_{++}> = \frac{\sqrt{2} N^{3/2}}{24 \pi} \lim_{z\rightarrow 0}
 \frac{1}{z^2} \left[\frac{\partial F}{\partial z} - z \nabla_{\bot}^2
F \right], 
\ee
\be
 <T_{++}> = \frac{N^2}{8\pi^2} \lim_{z \rightarrow 0} \frac{1}{z^3}
 \left[\frac{\partial F}{\partial z} - \frac{z}{2}
 \nabla_{\bot}^2 F \right],
\ee
\be
 <T_{++}> = \frac{N^3}{3 \pi^3} \lim_{z\rightarrow 0}
 \frac{1}{z^5} \left[\frac{\partial F}{\partial z} - \frac{z}{4}
 \nabla_{\bot}^2 F - \frac{z^3}{32}
 \left(\nabla_{\bot}^2\right)^2 F \right]
\ee
for $d=3,4,6$ respectively. 

The energy-momentum tensor is obviously dependent on the form
of the function $F$.  In particular, for the spurious wave with $F$ as
in (\ref{H}), we have
$<T_{++}>=0$, which is in agreement with the fact that our
spacetime in this case is just pure AdS in unusual coordinates.
Secondly, a Ricci-flat brane, with $F = F(x^+,x^i)$, also has 
$<T_{++}>=0$ since the Einstein equations in this case give 
$\nabla_{\bot}^2 F = 0$.  

In general, the function $F$ will have $z$-dependence that is a linear
combination of $z^d$ and $z^0$ as $z \rightarrow 0$. This can be seen
by Fourier analyzing equation (\ref{F_eqn}) with respect to $x^i$ to
obtain an equation in $z$ that can be solved with Bessel functions
\cite{CG}. Solutions proportional to $z^d$ give finite $<T_{++}>$
whereas those proportional to $z^0$ give a divergent result in
general. This is just the distinction between normalizable and
non-normalizable bulk modes \cite{BKL}, with the former corresponding
to a deformation of the state of the CFT and the latter to a change in
the parameters of the CFT Lagrangian, in this case the boundary metric.
We shall be interested only in the bulk solutions proportional to
$z^d$ at the boundary, for which the boundary is flat space and the
CFT is in a quantum state with non-vanishing $<T_{++}>$, i.e.,
a null momentum density. In fact, we shall restrict ourselves to
solutions of the form
\be
 F(x^+,x^i,z)= F(x^+) z^d.
\ee
The boundary energy-momentum tensor then describes a disturbance
propagating at the speed of light in the negative $x$-direction with wave
profile determined by $F(x^+)$. 

The case $F(x^+)={\rm constant}$ was studied in \cite{cvetic:99}. Starting
from a non-extremal stack of branes, it was shown that a finite boost
produces a metric describing a pp-wave propagating on the
branes. However, in order to retain the pp-wave in the extremal limit,
it is necessary to take the boost parameter to infinity. Therefore,
the metric (\ref{metric}) with $F(x^+,x^i,z)=Q/r^{\dt}$ ($Q={\rm
constant}$) can be regarded as describing an infinitely boosted stack
of BPS branes. Taking the decoupling limit keeping the momentum finite,
one obtains a sphere times an asymptotically AdS pp-wave spacetime with
\be
 F(x^+,x^i,z) = \mu^d z^d,
\ee
where $\mu$ is a constant.
In four dimensions, this pp-wave spacetime was first discussed by Kaigorodov
\cite{K1,K2}. Its $D$-dimensional generalization, denoted $K_D$, was
presented in \cite{cvetic:99}. The above discussion shows that $K_D$
can be regarded as an infinitely boosted version of AdS$_D$. The
infinite boost in the bulk induces an infinite boost on the conformal boundary,
and it is therefore natural to conjecture that supergravity theory in
the Kaigorodov spacetime (times a sphere of appropriate dimension) is
dual to a strongly coupled large $N$ CFT in an infinitely boosted
frame with constant background momentum density \cite{cvetic:99}. 

The Kaigorodov spacetime is a special case of the class of pp-wave
spacetimes that we have been considering, and we have seen that all of
these give a CFT energy-momentum tensor that describes a null momentum
density in the $x$-direction. In the case of $K_{d+1}$, the momentum density
is in fact constant: $P=<T_{++}>$ is given by
\be
 P = \frac{c \mu^2}{12 \pi},
\ee
\be
 P = \frac{\sqrt{2} N^{3/2} \mu^3}{8\pi},
\ee
\be
 P = \frac{N^2 \mu^4}{2 \pi^2},
\ee
\be
 P = \frac{2N^3 \mu^6}{\pi^3},
\ee
for $d=2,3,4,6$ respectively. Note that a {\it null} momentum density
is precisely what one would expect from an infinite boost. Our results
are therefore evidence in favour of the conjecture of \cite{cvetic:99}.
The classical supergravity approximation is valid in the large $N$ limit,
in which case $P \rightarrow \infty$. 

The three-dimensional Kaigorodov spacetime, $K_3$, is just that described by
the metric (\ref{ads3_wave}), with $F(x^+,z) = \mu^2 z^2$, and it is
curious that the energy-momentum tensor in this case
is non-vanishing, since we know from the above discussion that this spacetime
is just pure AdS$_3$ in unusual coordinates. The result can be
understood as follows~\cite{MS}. Start with pure AdS$_3$ in
horospherical coordinates, given by (\ref{ads_3}).  This corresponds
to the $SL(2,\mathbb{R})_L \otimes SL(2,\mathbb{R})_R$ vacuum of the
CFT, and the corresponding energy-momentum tensor will therefore vanish.
As we have seen, $K_3$ can be obtained from this metric via the
coordinate transformation (\ref{sl2r_u}), with
\be
F(x^+) = -\frac{1}{2}\{f,x^+\} = \mu^2,
\ee
a solution of which is
\be
f(x^+) = \frac{1}{2\mu} e^{2\mu x^+}.
\ee
The new boundary coordinates are ${x^+}' = f(x^+)$ and ${x^-}' =
x^-$. The boundary metric is 
\be
\label{eqn:boundary}
 ds^2 = dx^+ dx^- = e^{2\mu {x^+}'} {dx^+}' {dx^-}'.
\ee
The AdS/CFT calculation in $K_3$ gives the 
energy-momentum tensor in the conformal frame where the metric is
\be
 ds^2 = {dx^+}' {dx^-}',
\ee
which can be obtained from the metric (\ref{eqn:boundary}) by a Weyl
transformation. Under the Weyl transformation, the energy-momentum
tensor transforms as \cite{polchinski}\footnote{
We have inserted a factor of $-1/(2\pi)$ relative to the string theory
convention \cite{polchinski} in order to be consistent with
the usual Lorentzian field theory definition of the energy-momentum tensor.}:
\be
 T_{+'+'}' = T_{+'+'} - \frac{c}{24\pi} \{f,{x^+}'\},
\ee
Since $T_{+'+'}$ vanishes, this gives
\be
 T_{+'+'}' = \frac{c \mu^2}{12 \pi},
\ee
which is precisely what we obtained from our AdS/CFT calculation. 
Note that the Kaigorodov coordinates cover only the region $x^+>0$ of the
original spacetime. This is very similar to what occurs for the BTZ
black hole, and corresponds to adopting the coordinates of an
accelerating observer on the boundary. Such an observer would
naturally define the energy-momentum tensor so that it vanished in the
vacuum defined with respect to his time coordinate. This
energy-momentum tensor has a finite expectation value in the $SL(2,R)
\times SL(2,R)$ vacuum state and describes
the thermal radiation experienced by such an observer
\cite{MS}. 

Having understood the AdS$_3$ case, we now move on to discuss the
higher-dimensional generalizations of the Kaigorodov spacetime.

\sect{The Kaigorodov spacetime}

\subsection{The metric and its symmetries}

The $(d+1)$-dimensional Kaigorodov metric $K_{d+1}$ can be written
\be
 ds^2 = \frac{l^2}{z^2} \left( dx^+ dx^- + \mu^d z^d {dx^+}^2 + dx^i
 dx^i + dz^2 \right),
\ee
where $i = 1 \ldots d-2$. It will sometimes be useful to write
\be
 x^{\pm} = x \pm t.
\ee
The Kaigorodov spacetime is clearly asymptotic to
AdS$_{d+1}$ as $z \rightarrow 0$, and has a pp-curvature singularity
as $z \rightarrow \infty$. It was shown in~\cite{cvetic:99} that this
metric preserves $1/4$ supersymmetry, and has the following Killing
vector fields (we have converted the results of \cite{cvetic:99} to
our coordinate system)
\be
 K_+ = \sqrt{2} \frac{\partial}{\partial x^+}, \qquad K_- =
 \sqrt{2} \frac{\partial}{\partial x^-}, \qquad K_i =
 \frac{\partial}{\partial x^i},
\ee
\be
 L_{ij} = x^i \frac{\partial}{\partial x^j} - x^j
 \frac{\partial}{\partial x^i}, 
\ee
\be
 L_i = \frac{x^+}{\sqrt{2}} \frac{\partial}{\partial x^i} - \sqrt{2}
 x^i \frac{\partial}{\partial x^-},
\ee
\be
 J = -\frac{z}{l} \frac{\partial}{\partial z} - \frac{(d+2) x^-}{2l}
 \frac{\partial}{\partial x^-} + \frac{(d-2) x^+}{2l}
 \frac{\partial}{\partial x^+} - \frac{x^i}{l}
 \frac{\partial}{\partial x^i}.
\ee
The commutation relations of these Killing vector fields were
presented in \cite{cvetic:99}.

\subsection{Matching of symmetries}

The above Killing vector fields induce symmetries of the boundary
theory. In this subsection, we shall show that these symmetries are
the symmetries that one would expect for a CFT in a
background with the conformal symmetry broken by a constant (null)
momentum density in the $x^+$ direction. As we saw above, 
such a momentum density is described by an
energy-momentum tensor, the only non-vanishing component of which is
$P \equiv <T_{++}> = {\rm constant}$. 
The symmetries of the boundary theory are simply those conformal
transformations that leave this energy-momentum tensor invariant.

Some of the symmetries are obvious: the $K$ generators just give
boundary translations, and the $L_{ij}$ correspond to transverse
rotations. These transformations clearly leave the boundary metric
invariant, and preserve the form of the background energy-momentum
tensor. The other generators are more complicated. Let $y=x^1$ and
consider $L_y$. The finite form of the bulk isometry generated by this
Killing vector field is
\ba
\label{eqn:Litransfm}
 t' &=& \left(1 + \frac{\lambda^2}{4} \right) t + \frac{\lambda^2}{4} x
 + \frac{\lambda}{\sqrt{2}} y, \nonumber \\
 x' &=&  \left(1 - \frac{\lambda^2}{4} \right) x - \frac{\lambda^2}{4} t
 - \frac{\lambda}{\sqrt{2}} y \\
 y' &=& y + \frac{\lambda}{\sqrt{2}} t + \frac{\lambda}{\sqrt{2}} x. \nonumber
\ea
This is a combination of a boost in the $y$-direction, a rotation in
the $xy$ plane and a boost in the $x$-direction. Clearly the
parameters of the three transformations are related. To see how this
emerges, consider the transformations separately. A boost with
velocity $v$ in the $y$-direction followed by a rotation through angle
$\theta$ in the $xy$ plane gives
\be
\label{eqn:xplus}
 {x^+}' = \frac{1}{2} \left( \cos \theta + \gamma + \gamma v \sin \theta
 \right) x^+ + \frac{1}{2} \left( \cos \theta - \gamma - \gamma v \sin
 \theta \right) x^- - \gamma \left( \sin \theta + v\right) y,
\ee
where $\gamma = (1-v^2)^{-1/2}$. The energy momentum tensor after this
transformation becomes
\be
 < T_{\mu'\nu'} > = \frac{\partial x^+}{\partial {x^\mu}'} \frac{\partial
 x^+}{\partial {x^\nu}'} < T_{++} >,
\ee
so if the direction of momentum flow is to remain unchanged then we
require $x^+$ to be independent of ${x^-}'$ and $y'$. Inverting
equation \ref{eqn:xplus} (by replacing $v \rightarrow -v$ and $\theta
\rightarrow -\theta$), this requirement reduces to 
\be
 v = - \sin \theta, \qquad \cos \theta > 0.
\ee
The new energy momentum tensor then has only one non-vanishing
component, {\it viz}
\be
 <T_{+'+'}> = \cos^2 \theta <T_{++}>.
\ee
Finally, the $\cos^2 \theta$ prefactor can be removed by boosting
along the $x$-direction with velocity $w$, which gives
\be
 {x^+}'' = \sqrt{\frac{1-w}{1+w}} {x^+}',
\ee
so
\be
 < T_{+''+''} > = \frac{1+w}{1-w} \cos^2 \theta < T_{++} >,
\ee
hence if $w = \sin^2 \theta /(1+\cos ^2 \theta)$ then the energy
momentum tensor in the new frame is the same as in the original
frame. Therefore, the combined transformation is a symmetry of the
background of the boundary theory. In simple terms, it can be
understood as follows\footnote{
Although this is slightly misleading, since we are dealing with a
momentum {\it density}, which cannot be written as a four vector.}. 
In coordinates $t,x,x^i$, the momentum density
is $E(1,1,0,\ldots 0)$. (This is null because we are working in the
infinite momentum frame.) The boost in the $y$-direction takes this to
$E'(1, \cos \theta, -\sin \theta, \ldots 0)$ for some energy $E'$.
The rotation in the $xy$ plane brings this to $E' (1,1,0,\ldots 0)$. 
Finally, the boost in the $x$-direction brings this back to to $E (1,1,0
\ldots 0)$.  

The symmetries that we have discussed so far form the subgroup of the
Poincar\'e group that is unbroken by the background momentum
density. The remaining generator $J$ involves an unbroken conformal
symmetry. The finite bulk isometry induced by $J$ is
\be
\label{eqn:Jtransfm}
 {x^+}' = \lambda \lambda^{-d/2} x^+, \qquad {x^-}' = \lambda
 \lambda^{d/2} x^-,
 \qquad, {x^i}' = \lambda x^i, \qquad z' = \lambda z,
\ee
which is clearly a combination of a dilatation by factor $\lambda$,
and the transformation
\be
 x^{\pm} \rightarrow \lambda^{\mp d/2} x^{\pm}.
\ee
This latter transformation is simply a boost in the $x$-direction. To
see why this combination of dilatation and boost is a symmetry of the
boundary theory, consider replacing the boost by a general boost in
the $x$-direction with velocity $v$. For the boundary theory the
combined dilatation and boost is
\be
 {x^{\pm}}' = \lambda \sqrt{ \frac{1 \mp v}{1 \pm v} } x^+, \qquad {x^i}'
 = \lambda x^i.
\ee
The transformed energy-momentum tensor has one non-vanishing component: 
\be
 <T_{+'+'}> = \lambda^{-2} \frac{1+v}{1-v} <T_{++}>.
\ee
Clearly it is not invariant. However, neither is the flat boundary
metric:
\be
 ds^2 = \lambda^{-2} \left( {dx^+}' {dx^-}' + {dx^i}' {dx^i}' \right),
\ee
so to bring the metric back to the original form we must do a Weyl
transformation $g_{ij} \rightarrow \lambda^2 g_{ij}$. The Weyl
transformation of the energy-momentum tensor is
\be
 <\bar{T}_{+'+'}> = \lambda^{-(d-2)} <T_{+'+'}> = \lambda^{-d} \frac{1+v}{1-v}
 < T_{++}>.
\ee
The energy-momentum tensor is therefore invariant under the combined 
dilatation and boost provided we take $(1+v)/(1-v) = \lambda^d$, which
gives agreement with equation (\ref{eqn:Jtransfm}). 

Note that it is not find any combination of transformations that
involves special conformal transformations and leaves the
energy-momentum tensor invariant. 

\subsection{Scalar 2-point function}

It is well-known that conformal symmetry determines the 2-point
functions of operators up to an overall constant\footnote{
Since we are dealing with Lorentzian theories, an $i\epsilon$ term is
required, but suppressed throughout this paper.}:
\be
 \langle  {\cal O}(x) {\cal O}(x') \rangle =
 \frac{c_\Delta}{(x-x')^{2\Delta}},
\ee
where $\Delta$ is the scaling dimension of ${\cal O}$.
It is not surprising that AdS/CFT reproduces such 2-point functions
since they are entirely determined by symmetry, and AdS does indeed
give the correct conformal symmetry group.

We are interested in the case when some of the symmetry of the
boundary theory has been broken by a background momentum density. The
two point function in this case is less constrained. To see this,
write
\be
\langle {\cal O}(x_1^+,x_1^-,{\bf x}_1) {\cal O}(x_2^+,x_2^-,{\bf x}_2)
\rangle = f(x_{12}^+, x_{12}^-, {\bf x}_{12}^2 ),
\ee
where we have used the notation ${\bf x}$ for $x^i$, and $x_{12}$
denotes $x_1 - x_2$. In writing the two point function in this form,
we have made use of the unbroken translation and transverse rotation
symmetries. Now consider the unbroken symmetries generated by
$L_y$. Write ${\bf x} = (y,z,\ldots)$, and perform the transformation
(\ref{eqn:Litransfm}) with $\lambda = -\sqrt{2} y / x^+$. Invariance of
$f$ gives
\be
 f(x^+,x^-,y^2+z^2 + \ldots) = f(x^+, x^- + y^2/x^+,z^2 + \ldots),
\ee
and repeating the same procedure with $L_z$, etc. eventually gives
\be
 f(x^+, x^-, {\bf x}^2) = f(x^+, x^- + {\bf x}^2/x^+,1).
\ee
So the unbroken Poincar\'e symmetry has reduced the 2-point function
to a function of 2 variables, which can be taken to be $x^+$ and $x^2
= x^+ x^- + {\bf x}^2$. Let us denote this function by
$f(x^+,x^2)$. Now consider the final unbroken symmetry, generated by
$J$. If we assume that ${\cal O}$ has scaling dimension $\Delta$ then
under the dilatation plus boost symmetry, ${\cal O}$ transforms as
\be
 {\cal O}(x') = \lambda^{-\Delta} {\cal O}(x).
\ee
For the 2-point function, this implies that
\be
 f(\lambda^{-(d-2)/2} x^+, \lambda^2 x^2) = \lambda^{-2\Delta}
 f(x^+,x^2),
\ee
so choosing $\lambda^2 = 1/|x^2|$ gives
\be
 f(x^+,x^2) = \frac{1}{|x^2|^{\Delta}}
 f \left( x^+ |x^2|^{(d-2)/4},{\rm sign}(x^2) \right).
\ee
Hence the unbroken symmetries of the boundary theory restrict the
2-point function to take the form
\be
\label{eqn:2pointfn}
\langle {\cal O}(x_1^+,x_1^-,{\bf x}_1) {\cal O}(x_2^+,x_2^-,{\bf
 x}_2) \rangle  
 = \frac{1}{|x_{12}^2|^{\Delta}} f_{\pm} \left(
 x_{12}^+ |x_{12}^2|^{(d-2)/4} \right),
\ee
where $\pm = {\rm sign}(x^2)$. The argument of $f_{\pm}$ is invariant
under all the unbroken symmetries. In contrast with the conformally
invariant case, the functional dependence of the 2-point function is
not completely determined by symmetry. The AdS/CFT calculation of the
2-point function therefore has dynamical content for the Kaigorodov
spacetime, in contrast with the analogous calculation for pure AdS.

\subsection{AdS/CFT calculation of scalar 2-point function}

Having established that the scalar 2-point function is non-trivial, we
shall now calculate it using the AdS/CFT correspondence. The method
is similar to that of (\cite{freedman:99}) with the exception that we
can no longer Wick rotate to Euclidean signature since our the Kaigorodov
spacetime is not static. Lorentzian calculations for pure AdS were
discussed in \cite{BKL, BKLT}. We shall be careful
to examine the behaviour of surface terms that arise at the
singularity. 

Consider a scalar field of mass\footnote{We shall use units for
which $l=1$.} $m^2$: 
\be
 S_0 = \int d^{d+1}x \sqrt{-g} \left( -\frac{1}{2} (\partial \phi)^2 -
\frac{1}{2} m^2 \phi^2 \right).
\ee
The equation of motion is
\be
 (\nabla^2 - m^2) \phi = 0.
\ee
To compute the boundary 2-point function, we have to evaluate the
on-shell action of the scalar field solution $\phi(z,x)$ obeying
\be
 \phi(\epsilon,x) = \epsilon^{d-\Delta} \phi_0 (x),
\ee
where we have introduced a UV cut-off at $z=\epsilon$.
$\Delta$ is the scaling dimension of the
operator ${\cal O}$, related to the mass of $\phi$ by \cite{GKP,witten:98}
\be
 \Delta = \frac{d}{2} + \frac{1}{2}\sqrt{d^2+4m^2}.
\ee
The action can be integrated by parts and the equation of motion used,
giving
\be
 S_0 = \sum \int d^d x \sqrt{-h} \left( -\frac{1}{2} \phi n \cdot
\partial \phi \right),
\ee
where the sum runs over the two boundaries ($z=\epsilon$ and $z
\rightarrow \infty$). On each boundary, we have
outward unit normal $n = \mp (1/z) dz$ and the determinant of the
boundary metric is $(1/z)^d$. 

We shall solve for $\phi$ by Fourier analyzing the $x$ coordinates:
\be
 \phi(z,x) = \tilde{\phi}_k (z) e^{ik\cdot x},
\ee
where $k \cdot x = \frac{1}{2} (k^+ x^- + k^- x^+) + k^i x^i $. This
gives
\be
\label{eqn:phieq}
 z^{d-1} \frac{d}{dz} \left( \frac{1}{z^{d-1}} \frac{d
 \tilde{\phi}_k}{dz} \right) - \left( k^2 + \frac{m^2}{z^2} - (k^+)^2 \mu^d
 z^d \right) \tilde{\phi}_k = 0.
\ee
Note that when $k^+=0$, this equation is exactly the same as one would
obtain in pure AdS, with the solution \cite{freedman:99, BKLT}
\be
 \tilde{\phi}_k (z) = \epsilon^{d-\Delta} 
 \frac{z^{d/2} K_{\nu} \left(\sqrt{k^2} z
 \right)}{\epsilon^{d/2} K_{\nu} \left(\sqrt{k^2} \epsilon \right)}
 \tilde{\phi}_0(k),
\ee
where $\tilde{\phi}_0 (k)$ denotes the Fourier transform of $\phi_0 (x)$.
This solution is uniquely determined since the second linearly
independent solution gives rise to an infinite surface term at $z =
\infty$ \cite{freedman:99,BKLT}. 

Now consider the case $k^+ \ne 0$. 
For small $z$ (i.e. near infinity), neglecting the $z^d$ term gives
solutions in terms of Bessel functions. For $k^2 > 0$, one obtains 
\be
 \tilde{\phi}_k (z) = \epsilon^{d-\Delta} \frac{z^{d/2}}{\epsilon^{d/2}}
 \left(\frac{I_{-\nu} \left( \sqrt{k^2} z
 \right) + A(k) I_{\nu} \left( \sqrt{k^2} z \right)}{I_{-\nu} \left(
 \sqrt{k^2} \epsilon \right) + A(k) I_{\nu} \left( \sqrt{k^2} \epsilon
 \right)} \right)\tilde{\phi}_0 (k),
\ee
where $A(k)$ is some undetermined function and 
\be
 \nu = \frac{1}{2} \sqrt{d^2 + 4m^2},
\ee
which we assume is real, corresponding to the Breitenlohner-Freedman
bound \cite{breitenlohner:82} $m^2 > -d^2/4$. 
For $k^2 < 0$, one obtains
\be
 \tilde{\phi}_k (z) = \epsilon^{d-\Delta} \frac{z^{d/2}}{\epsilon^{d/2}}
 \left(\frac{J_{- \nu} \left( \sqrt{-k^2} z
 \right) + A(k) J_{\nu} \left( \sqrt{-k^2} z \right)}{J_{- \nu} \left(
 \sqrt{-k^2} \epsilon \right) + A(k) J_{\nu} \left( \sqrt{-k^2} \epsilon
 \right)} \right) \tilde{\phi}_0 (k).
\ee
For large $z$ (i.e. near the singularity), one can neglect the $k^2$
and $m^2/z^2$ terms and obtain the solutions\footnote{
In fact, the equation can also be solved exactly if one retains the
$m^2/z^2$ term and neglects only the $k^2$ term, giving the same solution
except that $\nu' = \sqrt{d^2 + 4m^2}/(d+2)$.}
\be
 \tilde{\phi}_k = z^{d/2} J_{\pm \nu'} \left( \frac{2 |k^+| \mu^{d/2}}{d+2}
 z^{(d+2)/2} \right),
\ee
where
\be
 \nu' = \frac{d}{d+2}.
\ee
Near the singularity, the solutions behave as
\be
 \tilde{\phi}_k \sim z^{(d-2)/4} \left\{ 
 \begin{array}{c}
  \sin  \\
  \cos   \\
 \end{array}
 \right\}
(z^{(d+2)/2}),
\ee
This implies that the surface
term near the singularity is bounded but oscillatory as 
$z \rightarrow \infty$. This is exactly what happens as $z \rightarrow
\infty$ for pure AdS when $k^2 < 0$ \cite{BKLT}, and can be dealt with
by introducing an infra-red cut-off. The two solutions
are equally acceptable near the singularity. In contrast with the behaviour in
pure AdS, it not possible to uniquely determine the solutions with
$k^2 >0$ by the requirement of finiteness of the surface term at $z =
\infty$. 

The interpretation of the arbitrary function $A(k)$ was explained in
\cite{BKLT}. This function corresponds to the freedom to choose
different Lorentzian propagators in the dual CFT. 
For example, one usually works with the Lorentzian 
Feynman propagator obtained by analytic
continuation of a unique bounded Euclidean propagator. However, there
are different propagators (for example the retarded propagator) which
one might instead choose to work with. The choice of propagator is
determined by boundary conditions, and the freedom to choose boundary 
conditions is encoded in the function $A(k)$. There may also be
freedom in the choice of vacuum state for the CFT.

In all cases, we have written the behaviour for small $z$ as 
\be
 \tilde{\phi}_k(z) = \epsilon^{d-\Delta} K(k,\epsilon,z)
 \tilde{\phi}_0(k),
\ee
where 
\be
 K(k,\epsilon,z) = 
 \frac{\tilde{\phi}^{-}_k(z) + A(k)
 \tilde{\phi}^{+}_k(z)}{\tilde{\phi}^{-}_k(\epsilon) + A(k)
 \tilde{\phi}^{+}_k(\epsilon)},
\ee
and $\tilde{\phi}^+$ and $\tilde{\phi}^-$ are the normalizable and 
non-normalizable \cite{BKL, BKLT} solutions, with the behaviour
\be
 \tilde{\phi}^+ \sim z^{\Delta}, \qquad \tilde{\phi}^- \sim
 z^{d-\Delta},
\ee
as $z \rightarrow 0$. Note that $A(k)$ is not arbitrary when $k^+=0$.
Evaluating the surface term at $z=\epsilon$
gives \cite{BKLT}
\be
 S_0 = \frac{1}{2} \int_{z=\epsilon} d^d k d^d k' \delta^d (k+k')
 \epsilon^{1-d} \tilde{\phi}_0(k) \partial_z K(k,\epsilon,z)
 \tilde{\phi}_0(k),
\ee
from which one can read off the 2-point function \cite{BKLT}
\be
 \langle {\cal O}(k) {\cal O}(k') \rangle = \delta^d (k+k')
 \lim_{\epsilon \rightarrow 0} \lim_{z
 \rightarrow \epsilon} \partial_z K(k,\epsilon,z).
\ee
At first sight, it appears that this 2-point function will be entirely
determined by the part of $K$ coming from the non-normalizable
solutions, since these terms dominate as $\epsilon \rightarrow 0$.
However, in pure AdS, this part of $K$ turns out to involve only
positive integer powers of $k^2$. 
After Fourier transforming back to position space, this gives
only contact terms (terms that vanish at non-zero
separation, e.g. derivative of delta functions). 
The non-trivial part of the 2-point function arises from
mixing between the normalizable and non-normalizable solutions, which
gives non-integer powers of $k^2$ \cite{freedman:99,BKLT}.

For the Kaigorodov spacetime, we
have to be a bit more careful since we know that $\tilde{\phi}^{\pm}$
will be corrected from their pure AdS forms by subleading terms that
depend on $\mu$. These subleading terms could dominate the terms
arising from mixing between the normalizable and non-normalizable
solutions, and therefore change the form of the 2-point function. We
can examine this $\mu$ dependence by seeking series solutions for
$\tilde{\phi}^{\pm}$ near $z=0$. One obtains\footnote{
If $\nu$ is an integer then there will be logarithmic terms in these
series. We shall
concentrate on the case of non-integral $\nu$ for simplicity.}:
\be
 \tilde{\phi}_k^- = z^{d-\Delta}\left(a_0 + a_2 z^2 + \ldots \right),
\ee
\be
 \tilde{\phi}_k^+ = z^{\Delta}\left(b_0 + b_2 z^2 + \ldots \right),
\ee
with the series agreeing with the approximate Bessel function
solutions up to $d$-th order, but then being altered by the $\mu$
dependence. The deviation from the Bessel function solutions at
$(d+2)$-th order is
\be
 \delta a_{d+2} = \frac{\mu^d {k^+}^2}{2(\Delta-d-1)(d+2)} a_0, \qquad
 \delta b_{d+2} = -\frac{\mu^d {k^+}^2}{2(\Delta+1)(d+2)} b_0.
\ee
Since the terms $a_0, \ldots a_d$, $b_0 \ldots b_d$ and the Bessel
function parts of $a_{d+2}$ and $b_{d+2}$ only involve
positive integer powers of $k^2$, it follows that $a_{d+2}$ and
$b_{d+2}$ only involve positive integer powers of $k^2$ and ${k^+}$,
which means that they will still only give contact terms in the
2-point function. Thus the non-trivial contribution to the 2-point 
function will come from the leading order terms in
$\tilde{\phi}^{\pm}$, which are independent of $\mu$. It follows that
the 2-point function will not depend on $\mu$. 

The only novel feature that arises from this calculation is that the
2-point function is uniquely determined in momentum space when $k^+ =
0$, but not when $k^+ \ne 0$. This is to be contrasted with the
calculation for pure AdS, which gives a unique propagator when $k^2 >
0$ but not when $k^2 < 0$, which is what one expects in field
theory\footnote{
We remind the reader why this is so. In free field theory, the
momentum space propagator is obtained from the equation
$(k^2+m^2)G(k;m^2) = 1$. The solution is uniquely defined except when $k^2
= -m^2$, corresponding to the freedom to add to $G(k;m^2)$ a function of
the form $g(k) \delta (k^2+m^2)$. In the interacting case, one can write
the 2-point function in a K\"allen-Lehmann representation as an
integral over free field propagators weighted by some spectral
function $\sigma (m^2)$. This spectral function vanishes for $m^2 <
0$, guaranteeing that the arbitrariness in the free field propagator
does not affect the interacting propagator when $k^2 > 0$.}.
The occurence of $k^+$ in our results is characteristic of the
infinite momentum frame. The freedom in the propagator when $k^+ \ne
0$ and $k^2 > 0$ is presumably related to some subtlety of
quantization in the infinite momentum frame.

Since there is no dependence on $\mu$, 
we are free to choose the function $A(k)$ so that the 2-point function
that we obtain from the Kaigorodov spacetime is the same as the
2-point function that we obtain from pure AdS by Wick rotation from
Euclidean signature. The only thing we have to
worry about it whether this gives the correct result when $k^+ = 0$,
which it must do since the bulk solution is the same for Kaigorodov 
and for pure AdS in this case. We therefore obtain the Feynman propagator:
\be
 \langle {\cal O}(x) {\cal O}(x') \rangle =
 \frac{c_{\Delta}}{((x-x')^2 - i\epsilon)^\Delta},
\ee
where $c_{\Delta}$ is given in \cite{freedman:99}. 

\subsection{Attempted explanation}

The simplicity of the AdS/CFT result suggests that there should be
some way of understanding it in purely field-theoretic terms. The best
we have come up with is the following. 

Writing the function $f$ of (\ref{eqn:2pointfn}) in terms of a
dimensionless variable, one obtains
\be
 f \left(P^{1/2} x^+ (x^2)^{(d+2)/4}, N, \lambda \right),
\ee
where we have now included the dependence on $N$ and $\lambda$
explicitly. Our AdS/CFT calculation is valid for $N,\lambda,P
\rightarrow \infty$ with $\mu^d \sim P/N^{d/2}$ fixed. We would like
to argue that $f$ must approach a constant in this limit, namely
$f(\infty,\infty,\infty)$. This would explain our AdS/CFT result. However,
this argument would break down if, at large $N$, $f$ reduced to a function of
\be
 N^{-d/4} P^{1/2} x^+ (x^2)^{(d-2)/4} \sim \mu^{d/2} x^+ (x^2)^{(d-2)/4}. 
\ee
In order to improve on this argument, one would need to understand the
explicit $N$-dependence of the 2-point function, at least for large $N$. 
If $P$ were kept finite in the large $N$ limit then perhaps it would
be possible to extend the usual planar-graph arguments (see
e.g. \cite{manohar:98}) of large $N$
field theory to fix the $N$-dependence of $f$. However, when $P$
scales with $N$, one is dealing with field theory in an $N$-dependent
background, which is probably much harder to understand.  

\sect{Discussion}

We have discussed spacetimes describing pp-waves propagating on the
worldvolume of extremal non-dilatonic branes. We have explicitly shown
that any inertial observer in the bulk of such a spacetime will
encounter infinite tidal forces at the place where the horizon would 
ordinarily be located.  This means that these solutions cannot be
extended through the horizon region, so the spacetime terminates at a
null singularity. This is evidence in support of the idea \cite{KMR,
HY} that there may be a generalized ``no-hair'' theorem that deals
with time-dependent ``hair''.

In the near-horizon limit, these spacetimes reduce to spacetimes
describing pp-waves moving along the horospheres of AdS. We have
discussed an apparent Virasoro symmetry \cite{BCG} of such spacetimes, 
and shown that it arises from an action of the Virasoro group on three
dimensional asymptotically AdS submanifolds. It would be interesting
to see whether this symmetry could be used to extend the entropy
calculation of \cite{strominger:98} to extended objects in higher
dimensional AdS, for example the BTZ black string of \cite{emparan:99b}.

We have shown that pp-waves moving tangential to the horospheres of
AdS will make the AdS horizon become singular unless the wave does not
extend as far as this horizon. This result has implications for the
Randall-Sundrum \cite{RS} scenario in which our universe is regarded
as a thin domain wall in AdS. This thin domain wall should presumably
be regarded as a mathematical idealization of a thick domain wall in AdS.
In \cite{CG}, it was shown that the pp-waves propagating tangentially to
such a domain wall correspond precisely to the gravitons of the Randall-Sundrum
scenario. We have shown that any of these pp-waves will induce a
curvature singularity at the AdS horizon. However, this horizon is the surface
at which one must impose boundary conditions for the bulk
spacetime. Thus one is faced with the unappealing prospect of having
to impose boundary conditions at a naked singularity, which presumably
requires an understanding of quantum gravity. If one were to impose
the boundary condition that the past horizon remain regular then it
appears that one must exclude the possibility of there being gravitons
present on the domain wall!

We have presented evidence in favour of the conjecture
\cite{cvetic:99} that gravity in the Kaigorodov spacetime is dual to a
CFT in the infinite momentum frame with constant background momentum
density. In particular, we have calculated the CFT energy momentum
tensor that arises from a large class of asymptotically AdS pp-waves,
including $K_{d+1}$. We have discussed how the isometries of $K_{d+1}$
have a natural interpretation as the subgroup of the conformal group
that leaves the background momentum density invariant. This symmetry
group determines the 2-point function of a scalar operator only up to an
arbitrary function of one variable. Our AdS/CFT calculation of the
2-point function therefore has dynamical content, and is found to
agree with the result one obtains from pure AdS (i.e., when the
conformal symmetry group is unbroken). We have suggested that this is
a large $N$ effect. The 2-point function exhibits some ambiguity
associated with working in Lorentzian signature, however the ambiguity
does not take the expected form. This is perhaps the only signal in
our work that the boundary CFT has really been quantized in the
infinite momentum frame.

It might be possible to understand our results more clearly by doing
a free field theory calculation in the CFT. The quantum state of the
CFT could be constructed by taking a thermal density matrix
(corresponding to a non-extremal stack of branes), applying a finite
boost, and then taking the limit in which the temperature tends to
zero and the boost to infinity with the energy density fixed. 

It would be interesting to have a complete understanding of how string
theory resolves the Kaigorodov singularity.  
Certainly, if we use the criterion
of Gubser \cite{steve} then this singularity is ``good'', in the sense that the
slightest finite temperature deformation will yield a non-extremal
brane-wave, which we know is simply a boosted non-extremal brane and
therefore has a regular horizon \cite{cvetic:99}.
On the other hand, it would be nice to know if the singularity in the bulk 
could be explicitly resolved along the lines of recent work 
\cite{jpp, ps} which deals primarily with timelike singularities.

We should also mention that our discussion of the Kaigorodov metric
has been confined to the case $\mu^d > 0$. However, the metric with
$\mu^d < 0$ is still a solution of Einstein's equations. This metric
gives a CFT energy-momentum tensor with negative energy, which is
unacceptable. So it appears that AdS/CFT is telling us that the $\mu^d
< 0$ Kaigorodov singularity cannot be resolved. Further support for
this idea comes from the fact that the metric in this case does not have a
regular finite temperature deformation.

As we have seen, it is easy to write down pp-wave solutions which are
asymptotic to AdS and which induce {\it non-constant} null momentum
density in the boundary theory.  For these spacetimes, we would
expect a non-trivial 2-point function in the CFT. It would be interesting
to investigate such spacetimes in more detail. A particular case of
interest would be to choose $F(x^+,z) = \mu^d {x^+}^p z^d$, with $p$
an integer (to avoid problems when $x^+<0$). Such a bulk spacetime
would admit a scaling symmetry analagous to the Kaigorodov scaling symmetry. 

Of course, not all gravitons in AdS will move tangential to the horospheres
(and therefore the boundary) of AdS.  As discussed in \cite{PST,lenny},
information about a graviton emitted from deep in the bulk of AdS which
heads towards the boundary of AdS must be encoded nonlocally in
Wilson loops of the boundary CFT, with
the size of the Wilson loop related to the distance of the disturbance
from the boundary.  These degrees of freedom in the CFT which encode 
information about the graviton in the bulk were dubbed ``precursors''.
An interesting (and open) problem is to find a full non-linear solution
describing a graviton moving in the direction transverse to the 
horospheres, so that we might have a hope of explicitly computing
properties of these precursors.

\bigskip

\centerline{\bf Acknowledgments}

DB would like to thank Simon Ross and Paul Saffin for discussions, and
Simon Ross for comments on a draft of this paper.  DB is supported in
part by the EPSRC grant GR/N34840/01. AC thanks Mirjam Cvetic, 
Dan Freedman and Andreas Karch for useful 
conversations.  AC is supported in part
by funds provided by the U.S. Department of Energy (D.O.E.) under
cooperative research agreement DE-FC02-94ER40818.
HSR thanks Jerome Gauntlett and Prem Kumar for discussions and is
funded by PPARC.

\appendix
\renewcommand{\theequation}
	     {\Alph{section}.\arabic{equation}}
\section{Appendix}
\setcounter{equation}{0}

Our line element is
\be
 ds^2 = H(r)^{-2/d} (2 dx^+ dx^- + F(x^+,x^i,r) dx^{+2} + dx^i dx^i) +
 H(r)^{2/\dt} (dr^2 + r^2 d\Omega_{\dt+1}^2).
\ee
We shall compute the curvature tensors in the following basis:
\ba
 e^+ & = & H^{-1/d} dx^+, \qquad e^- = H^{-1/d} \left(dx^- + \frac{1}{2} F
 dx^+ \right), \qquad e^{i} = H^{-1/d} dx^i, \nonumber \\
 \qquad e^r & = & H^{1/\dt} dr, \qquad e^m = H^{1/\dt} r \hat{e}^m,
\ea
where $\hat{e}^m$ are an orthonormal basis of 1-forms with respect to
the metric $\hat{g}_{mn}$ associated with the line element
$d\Omega_{\dt+1}^2$. The metric can be written
\be
 g_{MN} = \eta_{AB} e^{A}_{\;\; M} e^{B}_{\;\; N},
\ee
where
\ba
 \eta_{+-} = \eta_{-+} & = & \eta_{rr} = 1 \nonumber \\
             \eta_{ij} & = & \delta_{ij} \\
             \eta_{mn} & = & \delta_{mn} \nonumber
\ea
is the Minkowski metric. Note that we are using letters $M,N,\ldots$
as curved (coordinate) indices and $A,B,\ldots$ as basis indices.

The connection 1-forms are defined by
\be
 de^{A} = -\omega^A_{\;\; B} \wedge e^B.
\ee
The non-vanishing connection 1-forms are
\be
 \omega_{+i} = \omega^-_{\; \; i}= \frac{1}{2} H^{1/d} \frac{\partial F}
 {\partial x^i} e^+,
\ee
\be
 \omega_{+r} = \omega^-_{\; \; r}  = -\frac{1}{d} H^{-1/\dt} \frac{H'}{H} e^- +
 \frac{1}{2} H^{-1/\dt} \frac{\partial F}{\partial r} e^+,
\ee
\be
 \omega_{-r} = \omega^+_{\; \; r} = -\frac{1}{d} H^{-1/\dt} \frac{H'}{H} e^+,
\ee
\be
 \omega_{ir} = \omega^i_{\; \; r} = -\frac{1}{d} H^{-1/\dt} \frac{H'}{H} e^i,
\ee
\be
 \omega_{rm} = \omega^r_{\; \; m} = -H^{-1/\dt} \left(\frac{H'}{\dt H} +
 \frac{1}{r} \right) e^m,
\ee
\be
 \omega_{mn} = \hat{\omega}_{mn},
\ee
where $\hat{\omega}_{mn}$ are the connection 1-forms associated with
$\hat{e}^m$. Other non-vanishing components are obtained from the ones
given above by antisymmetry: $\omega_{AB} = -\omega_{BA}$. 

The curvature 2-form is defined by
\be
 \Theta_{AB} = d\omega_{AB} + \omega_{AC} \wedge \omega^C_{\;\; B}.
\ee
The non-vanishing curvature 2-forms are
\be
 \Theta_{+-} = \frac{1}{d^2} H^{-2/\dt} \left(\frac{H'}{H}\right)^2 e^+
 \wedge e^-,
\ee
\ba
 \Theta_{+i}  & = & \frac{1}{2d} H^{-2/\dt} \frac{H'}{H} \frac{\partial
 F}{\partial r} e^+ \wedge e^i - \frac{1}{2} H^{2/d} \frac{\partial^2
 F}{\partial x^i \partial x^j} e^+ \wedge e^j \nonumber \\
 & - & \frac{1}{2} H^{1/d-1/\dt} \frac{\partial^2 F}{\partial x^i
 \partial r} e^+ \wedge e^r - \frac{1}{d^2} H^{-2/dt}
 \left(\frac{H'}{H}\right)^2 e^- \wedge e^i,
\ea
\ba
 \Theta_{+r} & = & -\frac{1}{2} H^{1/d-1/\dt} \frac{\partial^2 F}
 {\partial x^i \partial r} e^+ \wedge e^i + \frac{1}{2} H^{-2/\dt} 
 \left( -\frac{\partial^2 F}{\partial r^2} + \left(\frac{2}{d} +
 \frac{1}{\dt} \right) \frac{H'}{H} \frac{\partial F}{\partial r}
 \right) e^+ \wedge e^r \nonumber \\ &+& \frac{1}{d} H^{-2/\dt} \left(
 \frac{H''}{H} - \left(\frac{1}{d} + \frac{1}{\dt} + 1\right)
 \left(\frac{H'}{H} \right)^2 \right) e^- \wedge e^r,
\ea
\be
 \Theta_{+m} = -\frac{1}{2} H^{-2/\dt} \left(\frac{H'}{\dt H} +
 \frac{1}{r} \right) \frac{\partial F}{\partial r} e^+ \wedge e^m +
 \frac{1}{d} H^{-2/\dt} \frac{H'}{H} \left(\frac{H'}{\dt H} +
 \frac{1}{r} \right) e^- \wedge e^m,
\ee
\be
 \Theta_{-i} = -\frac{1}{d^2} H^{-2/\dt} \left(\frac{H'}{H}\right)^2
 e^+ \wedge e^i,
\ee
\be
 \Theta_{-r} = \frac{1}{d} H^{-2/d} \left( \frac{H''}{H} - \left(
 \frac{1}{d} + \frac{1}{\dt} + 1\right) \left(\frac{H'}{H}\right)^2
 \right) e^+ \wedge e^r,
\ee
\be
 \Theta_{-m} = \frac{1}{d} H^{-2/\dt} \frac{H'}{H} \left(\frac{H'}{\dt
 H} + \frac{1}{r} \right) e^+ \wedge e^m,
\ee
\be
 \Theta_{ij} = -\frac{1}{d^2} H^{-2/\dt} \left(\frac{H'}{H}\right)^2
 e^i \wedge e^j,
\ee
\be
 \Theta_{ir} = \frac{1}{d} H^{-2/\dt} \left(\frac{H''}{H} - \left(
 \frac{1}{d} + \frac{1}{\dt} + 1\right) \left(\frac{H'}{H}\right)^2
 \right) e^i \wedge e^r,
\ee
\be
 \Theta_{im} = \frac{1}{d} H^{-2/\dt} \frac{H'}{H} \left(\frac{H'}{\dt
 H} + \frac{1}{r}\right) e^i \wedge e^m,
\ee
\be
 \Theta_{rm} = -\frac{1}{\dt} H^{-2/\dt} \left(\frac{H''}{H} -
 \left(\frac{H'}{H}\right)^2 + \frac{H'}{Hr}\right) e^r \wedge e^m,
\ee
\be
 \Theta_{mn} = \hat{\Theta}_{mn} - H^{-2/\dt} \left(\frac{H'}{\dt H} +
 \frac{1}{r} \right)^2 e^m \wedge e^n, 
\ee
with the other non-vanishing components related to these by
antisymmetry. The Riemann curvature tensor is defined by
\be
 \Theta_{AB} = \frac{1}{2} R_{ABCD} e^C \wedge e^D.
\ee
The non-zero components are
\be
 R_{+-+-} = \frac{1}{d^2} H^{-2/\dt} \left(\frac{H'}{H}\right)^2,
\ee
\be
 R_{+i+j} = -\frac{1}{2} H^{2/d} \frac{\partial^2 F}{\partial x^i
 \partial x^j} + \frac{1}{2d} H^{-2/\dt} \frac{H'}{H} F' \delta_{ij},
\ee
\be
 R_{+i+r} = -\frac{1}{2} H^{1/d-1/\dt} \frac{\partial^2 F}{\partial r
 \partial x^i},
\ee
\be
 R_{+i-j} = -\frac{1}{d^2} H^{-2/\dt} \left(\frac{H'}{H}\right)^2
 \delta_{ij},
\ee
\be
 R_{+r+r} = \frac{1}{2} H^{-2/\dt} \left( -\frac{\partial^2
 F}{\partial r^2} + \left(\frac{2}{d}+\frac{1}{\dt}\right)
 \frac{H'}{H} \frac{\partial F}{\partial r} \right),
\ee
\be
 R_{+r-r} = \frac{1}{d} H^{-2/\dt} \left( \frac{H''}{H} -
 \left(\frac{1}{d} + \frac{1}{\dt} +
 1\right)\left(\frac{H'}{H}\right)^2 \right),
\ee
\be
 R_{+m+n} = -\frac{1}{2} H^{-2/\dt} \left( \frac{H'}{\dt H} +
 \frac{1}{r} \right) F' \delta_{mn},
\ee
\be
 R_{+m-n} = \frac{1}{d} H^{-2/\dt} \frac{H'}{H} \left( \frac{H'}{\dt H} +
 \frac{1}{r} \right) \delta_{mn},
\ee
\be
 R_{ijkl} = -\frac{1}{d^2} H^{-2/\dt} \left(\frac{H'}{H}\right)^2
 (\delta_{ik} \delta_{jl} - \delta_{il} \delta_{jk}),
\ee
\be
 R_{irjr} = \frac{1}{d} H^{-2/\dt} \left( \frac{H''}{H} -
 \left(\frac{1}{d} + \frac{1}{\dt} +
 1\right)\left(\frac{H'}{H}\right)^2 \right) \delta_{ij},
\ee
\be
 R_{imjn} = \frac{1}{d} H^{-2/\dt} \frac{H'}{H} \left( \frac{H'}{\dt H} +
 \frac{1}{r} \right) \delta_{ij} \delta_{mn},
\ee
\be
 R_{rmrn} = -\frac{1}{\dt} H^{-2/\dt} \left(\frac{H''}{H} -
 \left(\frac{H'}{H} \right)^2 + \frac{H'}{Hr} \right) \delta_{mn},
\ee 
\be
 R_{mnpq} = \hat{R}_{mnpq} - H^{-2/\dt} \left(\frac{H'}{\dt H} +
 \frac{1}{r} \right)^2 (\delta_{mp} \delta_{nq} - \delta_{mq}
 \delta_{np}),
\ee
Other non-vanishing components are related to these by the symmetries
of the Riemann tensor. The Ricci tensor is
\be
 R_{AB} = R_{A \; \; BC}^{\; \; C} = R_{A-B+} + R_{A+B-} + R_{AiBi} +
 R_{ArBr} + R_{AmBm}.
\ee
The non-vanishing components are
\be
 R_{++} = -\frac{1}{2} H^{-2/\dt} \left(\frac{\partial^2 F}{\partial
 r^2} + \frac{\dt+1}{r} \frac{\partial F}{\partial r} \right) -
 \frac{1}{2} H^{2/d} \nabla_{\bot}^2 F,
\ee
where $\nabla_{\bot}^2$ is the Laplacian with respect to the flat
coordinates $x^i$,
\be
 R_{+-} = \frac{1}{d} H^{-2/\dt} \left(\frac{H''}{H} -
 \left(\frac{H'}{H} \right)^2 + \frac{\dt + 1}{r} \frac{H'}{H}\right),
\ee
\be
 R_{ij} = \frac{1}{d} H^{-2/\dt} \left(\frac{H''}{H} -
 \left(\frac{H'}{H} \right)^2 + \frac{\dt + 1}{r} \frac{H'}{H}\right)
 \delta_{ij},
\ee
\be
 R_{rr} = -\frac{1}{\dt} H^{-2/\dt} \left( \frac{H''}{H} +
 \frac{\dt}{d} \left(\frac{H'}{H}\right)^2 + \frac{\dt+1}{r}
 \frac{H'}{H} \right),
\ee
\ba
 R_{mn} &=& \hat{R}_{mn} - \frac{1}{\dt} H^{-2/\dt} \left(\frac{H''}{H} -
 \left(\frac{H'}{H} \right)^2 + \frac{\dt + 1}{r} \frac{H'}{H} +
 \frac{1}{r^2} \right) \delta_{mn} \nonumber \\
        &=& -\frac{1}{\dt} H^{-2/\dt} \left(\frac{H''}{H} -
 \left(\frac{H'}{H} \right)^2 + \frac{\dt + 1}{r} \frac{H'}{H}\right)
 \delta_{mn},
\ea
where in the final expression we have used the fact that the
`internal' space is an Einstein space, which in our basis implies
\be
 \hat{R}_{mn} = \frac{\dt}{r^2} H^{-2/\dt} \delta_{mn}.
\ee

\end{document}